\RequirePackage{fix-cm}
\documentclass{svjour3}                     
\smartqed
\usepackage{graphicx}
\usepackage{mathptmx}      
\usepackage[colorlinks,citecolor=blue]{hyperref}
\usepackage[T1]{fontenc}
\usepackage{cite}
\usepackage{amsmath}
\usepackage{amsfonts}
\usepackage{array}
\usepackage[caption=false,font=footnotesize]{subfig}
\usepackage{amsmath}
\usepackage{xfrac}
\usepackage{color}
\usepackage[ruled]{algorithm2e}
\SetAlgorithmName{Table}{Table}

\newcommand{\e}{\textrm{e}}
\newcommand{\nakagami}{Nakagami-\textit{m}~}
\newcommand{\h}{\mathsf{h}}

\journalname{Wireless Networks} 

\begin{document}	
	\title{Transmission Delay Minimization in Wireless Powered Communication Systems}		
	\titlerunning{Transmission Delay Minimization in Wireless Powered Communication Systems}
	
	\author{Mohammad Lari}
	\institute{M. Lari \at
		Electrical and Computer Engineering Faculty, Semnan University, Semnan, Iran \\
		\email{m\_lari@semnan.ac.ir}}
	
	\date{Received: date / Accepted: date}	
	\maketitle
	
	\begin{abstract}
		We study transmission delay minimization of a wireless powered communication (WPC) system in a point-to-point scenario with one hybrid access point (HAP) and one WPC node. In this type of communications, the HAP sends energy to the node at the downlink (DL) for a given time duration and the WPC node harvests enough radio frequency (RF) power. Then, at the uplink (UL) channel, the WPC node transmits its collected data in a given time duration to the HAP. Minimizing such round trip delay is our concern here. So, we have defined four optimization problems to minimize this delay by applying the optimal DL and UL time durations and also the optimal power at the HAP. These optimization problems are investigated here with thorough comparison of the obtained results. After that, we extend our study to the multiuser case with one HAP and $K$ nodes and two different optimization problems are studied again in these cases.	
		
	\end{abstract}


\section{Introduction}\label{sec:introduction}
With the impressive growth of electronic devices and their diverse usage in current days, provision of sufficient power for these devices has become a challenging issue. For example, in a wireless sensor network with multiple and scattered nodes, usually it is not possible to connect all the nodes to the power grid (electricity grid), so, the nodes have to use battery. Therefore, replacement or recharging of the batteries in a specific time duration is necessary. This task is not always implemented easily. For instance, doing this for such sensors which are installed in an impassable area or  mobile devices or some military equipment are always difficult and even for the implanted sensors at the human bodies is almost impossible. Consequently, for such devices, an accurate mechanism or proper circuits are embedded to recharge the battery by using the available environmental energy \cite{Surveys-Tut-1}. 

Energy harvesting from nature is a wide topic, and much research has been directed in this context yet \cite{Surveys-Tut-3}. Energy harvesting method typically uses the solar and wind energy or ambient radio frequency (RF) radiation. Despite the advantages of these types of energy, their usage is not possible at all times (i.e. solar energy is not accessible at night or RF radiation depends on the radiators activity). Therefore, in a modern way and with a well managed method, the energy is generated in a place and transmitted to another place for its end-user. Normally, this type of energy is generated in the form of an electromagnetic wave and transmitted wirelessly to its consumers \cite{Wireless-Comm-Mag-1,MIMO-1}. 

Due to nature of the wireless channel, power attenuation in wireless power transfer (WPT) are very high, and the efficiency of WPT is very low, especially over long distance. But, on the other hand, WPT administrator has full control over its power transfer, where the transmit power, waveforms, occupied time/frequency dimensions and so on, are all tunable for providing stable energy supply under different physical conditions and service requirements. Thanks to this evident advantages of WPT over conventional method, WPT obtains much interest, and even some products are manufactured based on this technology recently \cite{Surveys-Tut-1}. In addition, much research is being conducted to improve the efficiency of WPT which would lead to the vast deployment of this technique in the near future \cite{Wireless-Comm-Mag-2,Microwave-Mag-1}. In this regard, integration of wireless-based energy harvesting in wireless communication systems becomes a new interesting challenge recently \cite{Comm-Mag-1}. In this way, the harvested energy can help low power nodes or sensors to transfer their own data more efficiently to the destinations.

There are two major directions for implementing wireless powered communication (WPC) system which wirelessly receives the power and wirelessly transmits data: one is referred to as the simultaneous wireless information and power transfer (SWIPT) system \cite{SWIPT-Main,SWIPT-1,SWIPT-2,SWIPT-3,SWIPT-4,SWIPT-5,Zahedi,SWIPT-6,SWIPT-7,SWIPT-8,SWIPT-9,SWIPT-10}, and the other transfer power and information separately \cite{MIMO-1,MIMO-2,MIMO-3,MIMO-4,MIMO-5,Resource-Allocation-4,Resource-Allocation-5,Rui-Zhang-1,Rui-Zhang-2,Resource-Allocation-6,Main-2016-1,Simple-1,Simple-2,Dedicated-1,Dedicated-2,Resource-Allocation-1,Resource-Allocation-2,X-1,X-2}. The second scheme is more simple for practical implementation and will be studied in our paper in the following. So, here we assume a hybrid access point (HAP) which transfers power to the nodes in the downlink (DL) channel and then, after sufficient energy absorption by the nodes, wireless sensors transmit their own data in the uplink (UL) channel to the HAP \footnote{The HAP may have some information for the nodes and can transmit them in the DL in different time slot. However, without lose of generality, information transmission from the HAP to the nodes is not our concern here.}. Typically, this protocol is termed "harvest-then-transmit" in literature.

In some applications such as tactile internet, factory automation and vehicle collision avoidance, the overall packet size is small. However, these short packets require very small latency as their quality-of-service (QoS) \cite{Tactile-Internet}. Under these conditions, optimization of WPC to minimize the total DL/UL time duration is more desirable. Note that, more time for harvesting energy at the DL phase leads to higher available transmit power at the UL phase. Again, higher available power at the UL phase leads to higher data rate and lower time duration at the UL phase. As the same way, less time for harvesting energy at the DL phase leads to higher time duration at the UL phase as well. Consequently, finding the optimal (minimum) values for the DL/UL time duration is our target here. Such kind of optimization for minimizing the transmission delay (TD) from the HAP to the node in the DL and from the node to the HAP in the UL is studied here within four different problems. In each optimization problems, we attempt to adjust the DL/UL time duration with or without power allocation to minimize the total delay in the \nakagami flat fading channel. Since \nakagami model approximates some popular channel models such as Rician (see \cite{Dedicated-1,Dedicated-2}), we use \nakagami for the small scale fading model in our channel \cite{Alouini}. After that, we extend the TD minimization problem to the multiuser case with one HAP and $K$ nodes \cite{Recommended}. In the multiuser scenario, the HAP first transmits power to the nodes at the DL and then, the nodes transmit data to the HAP in different time one by one in a time division multiple access (TDMA) manner. Then, minimizing the TD with and without power allocation at the HAP is studied in two different problems. The main contributions of this paper is summarized as follows:
\begin{itemize}
\item
New optimization problems which minimize the TD instead of the throughput rate is studied here. Delay minimization is more attractive in some applications require low latency. To the best of our knowledge, this problem is not investigated in WPC widely.
\item
In single user scenario, we define four optimization problems $\mathcal{P}_1-\mathcal{P}_4$, which are distinct from the previous problems in literature and we solve them here.
\\
$\mathcal{P}_1:$ TD minimization without the optimal power allocation when the DL and UL time duration are equal. 
\\
$\mathcal{P}_2:$ TD minimization with the optimal power allocation when the DL and UL time duration are equal.
\\
$\mathcal{P}_3:$ TD minimization without the optimal power allocation when the DL and UL time duration are not necessarily equal.
\\
$\mathcal{P}_4:$ TD minimization with the optimal power allocation when the DL and UL time duration are not necessarily equal. In the problem $\mathcal{P}_4$, derivation of the optimal solution for the allocated power is so complicated even numerically. Therefore, a very tight approximation is employed for finding the solution in this case.
\item
In multiuser scenario, we define two optimization problems $\mathcal{P}_5-\mathcal{P}_6$ for one HAP and $K$ nodes and solve them numerically.
\\
$\mathcal{P}_5:$ TD minimization without the optimal power allocation when the sum of $K$ UL time slots are less than or equal to the DL time slot.
\\
$\mathcal{P}_6:$ TD minimization with the optimal power allocation when the sum of $K$ UL time slots are less than or equal to the DL time slot.
\end{itemize}

The rest of our paper is organized as follows. First, some related work is reviewed in Section \ref{sec:related_work} and  the system model is introduced in Section \ref{sec:system_model} where a WPC system with one HAP and one WPC node is discussed. In Section \ref{sec:rtt_min}, four optimization problems for minimization of the total delay in a WPC system are considered completely. Then, in Section \ref{sec:extension_multiuser}, TD minimization of Section \ref{sec:rtt_min} is extended and investigated in a multiuser case. At last, the simulation results are presented in Section \ref{sec:simulation_results}, and Section \ref{sec:conclusion} concludes the paper.

For ease of reading, the abbreviations used in this article are summarized in Table \ref{tab:1} alphabetically. In addition, we use $\mathbb{E}\{.\}$ for the statistical expectation, $\Gamma(.)$ for the gamma function \cite{Ryzhik} and $\mathcal{W}_0(.)$ for the Lambert-W function \cite{lambertw}.

\begin{table}
  \centering
  \caption{List of abbreviations.}\label{tab:1}
  \begin{tabular}{|c|c|}
  \hline
  \textbf{Abbreviation} & \textbf{Description} \\
  \hline\hline
  BER & Bit error rate \\
  \hline
    CSI & Channel state information \\
  \hline
    DL/UL & Downlink/Uplink \\
  \hline
    HAP & Hybrid access point \\
  \hline
  HD/FD & Half-duplex/Full-duplex \\
  \hline
  MIMO & Multiple-input multiple-output \\
  \hline
  PB & Power beacon \\
  \hline 
  PDF & Probability density function \\
  \hline 
  QoS & Quality-of-service \\
  \hline
  RF & Radio frequency \\
  \hline
  RFID & Radio frequency identification \\
  \hline  
  SWIPT & Simultaneous wireless information and power transfer \\
  \hline
  TD & Transmission delay \\
  \hline
  TDMA & Time division multiple access \\
  \hline
  WPC & Wireless powered Communication \\
  \hline
  WPT & Wireless power transfer \\
  \hline  
\end{tabular}
\end{table}


\section{Related Work}\label{sec:related_work}
There are two major configurations for wireless power and information transfer \footnote{A complete survey can be found in \cite{Surveys-Tut-3,Surveys-Tut-2}.}. In the first configuration, wireless power transfer and information transmission carry out simultaneously and this method abbreviated as SWIPT. Number of studies consider SWIPT in various structure and context such as \cite{SWIPT-1,SWIPT-2,SWIPT-3} in point-to-point and broadcast channels, \cite{SWIPT-4,SWIPT-5,Zahedi} in the relay based systems, \cite{Zahedi,SWIPT-6,SWIPT-7} in multiple-input multiple-output (MIMO) channels, \cite{SWIPT-8} for opportunistic channels and \cite{SWIPT-9,SWIPT-10} in the cooperative systems. On the other hand, the second configuration focuses on separated power and information transfer specially in time domain and it is well-known as "harvest-then-transmit" in the literature. Harvest-then-transmit protocol is more simpler than the SWIPT and we formulate our problem within this protocol later.

Harvest-then-transmit protocol is studied in many research. For example \cite{MIMO-1,MIMO-2,MIMO-3,MIMO-4,MIMO-5} investigated this scheme with multiple antenna systems. In \cite{MIMO-1}, TDMA users are served by a multiple antenna HAP and sum throughput rate of the WPC network is maximized with joint energy beamforming and time allocation at the DL and UL links. In \cite{MIMO-2}, the authors assumed a large-scale MIMO system and maximized energy efficiency of information with energy beamforming in the system. The authors of \cite{MIMO-3} focused on the tradeoff of wireless energy and information transfer by adjusting the transfer duration with a total duration constraint and derived two wireless energy and information transfer tradeoff schemes by maximizing an upper bound and an approximate lower bound of the average information transmission rate, respectively. In \cite{MIMO-5}, the impact of channel state information (CSI) and antenna correlation at the multi-antenna  wireless powered relay is investigated. Tow different scenarios for the availability of the CSI are assumed and the analytical expressions for the outage probability and ergodic capacity are derived. Moreover, maximizing the minimum rate among all users in the massive MIMO with an asymptotic optimal solution is obtained in \cite{Resource-Allocation-4}. Some interesting insights for the optimal design of massive MIMO WPC system are also included. As the same way, in \cite{Resource-Allocation-5} with using the stochastic geometry, the spatial throughput of wireless nodes is maximized by jointly optimizing frame partition between DL and UL phases. The impact of battery storage on the spatial throughput is also illustrated.

Resource allocation to maximize the sum throughput rate is studied completely by Ju \textit{et al.} within a half-duplex (HD) \cite{Rui-Zhang-1} and full-duplex (FD) network \cite{Rui-Zhang-2} respectively. In \cite{Rui-Zhang-1}, a HAP transmits power for the users at the DL and after sufficient energy harvesting, users transmit information to the HAP in different time slots. Weighted sum throughput of the network is considered and a closed-form solution for the optimal DL time slot and UL time slots is derived. In addition, a doubly near-far problem is introduced and the authors presented a new solution to resolve the problem. However, the solution may reduce the sum throughput rate of the network. The doubly near-far problem is also discussed in \cite{MIMO-4} and some solutions by using proper beamforming technique is designed. Then \cite{Rui-Zhang-2} extended \cite{Rui-Zhang-1} with the FD HAP to improve the overall performance. The optimal and suboptimal solutions for the allocated time slots at the DL and UL is obtained under perfect and imperfect self-interference cancellation. In particular, in \cite{Resource-Allocation-6}, the performance of WPC in the presence of statistical queuing constraints with buffer overflow probability \cite{Lari} is studied. Then, the optimal time allocation for energy harvesting and information decoding operations depends on these constraints is obtained. 

Different from \cite{Rui-Zhang-1} and \cite{Rui-Zhang-2}, references \cite{Main-2016-1,Simple-1,Simple-2,Dedicated-1,Dedicated-2} assumed two dedicated energy and data access points. The authors of \cite{Main-2016-1} intended to maximize the achievable throughput by balancing the time duration of wireless power transfer at the DL and the information transfer at the UL while satisfying the energy causality, time duration and QoS constraints. Similar to \cite{Main-2016-1}, references \cite{Simple-1,Simple-2} assumed dedicated access points for the energy and data separately to maximize the achievable rate as well. In \cite{Simple-1}, the devices opportunistically access the wireless charging channel and information transmission channel and a power control strategy to minimize the energy consumption was presented. A wirless power beacon (PB) which performs channel estimation, digital beamforming and spectrum sensing is assumed in \cite{Dedicated-1} where with a time-splitting approach, the source node first harvest energy from the PB and then, transmits information to the destination. Analytical experssions for the average throughput are derived and the optimal time split to maximize that is extracted. In addition, the impact of cochannel interference is studied. Finally, some notable points about the number of antennas and the role of cochannel interference are shown. In \cite{Dedicated-2}, the authors present the ergodic capacity of a similar system model to \cite{Dedicated-1}. The energy transfer link is subjected to Rician fading, which is a real fading environment due to relatively short range power transfer distance and the existence of a strong line of sight path. 

Similar to harvest-then-transmit protocols, harvest-then-cooperate protocol can be used in a relay based communication to improve network performance \cite{Resource-Allocation-1,Resource-Allocation-2,X-1,X-2,Delay-Aware}. In this protocol, the source and relay first harvest energy and then transmit source's information to the destination cooperatively.
In \cite{Resource-Allocation-1} an approximate closed-form expression for the average throughput of a WPC cooperative network over Rayleigh fading channels is derived. Again, in \cite{Resource-Allocation-2}, for the cooperative scenario, a social welfare maximization problem to maximize the weighted sum throughput of all HAP-source pairs, which is subsequently solved by a water-filling based distributed algorithm is formulated. The optimal energy beamforming vector and the time split between harvest and cooperation are investigated in \cite{X-1}. Then, a closed-form expressions for the energy beamforming vector and the time split are extracted. 

As we can see, the most previous articles have focused on optimization of the throughput rate. Nevertheless, in many cases such as tactile internet, factory automation and vehicle collision avoidance \cite{Tactile-Internet}, data transmission with the low latency is more desirable. Under these conditions, optimization of the total DL/UL time duration instead of throughput rate is adopted. For this reason, transmission delay minimization in harvest-then-transmit WPC system is considered in our paper. Although our system model is similar to the previously mentioned papers (especially \cite{Rui-Zhang-1},\cite{Rui-Zhang-2} and \cite{Main-2016-1}), but, the criterion and the cost function of our optimization problem are totally different. 


\section{System Model}\label{sec:system_model}
\begin{figure}
	\begin{center}
		\includegraphics[draft=false,width=0.95\linewidth]{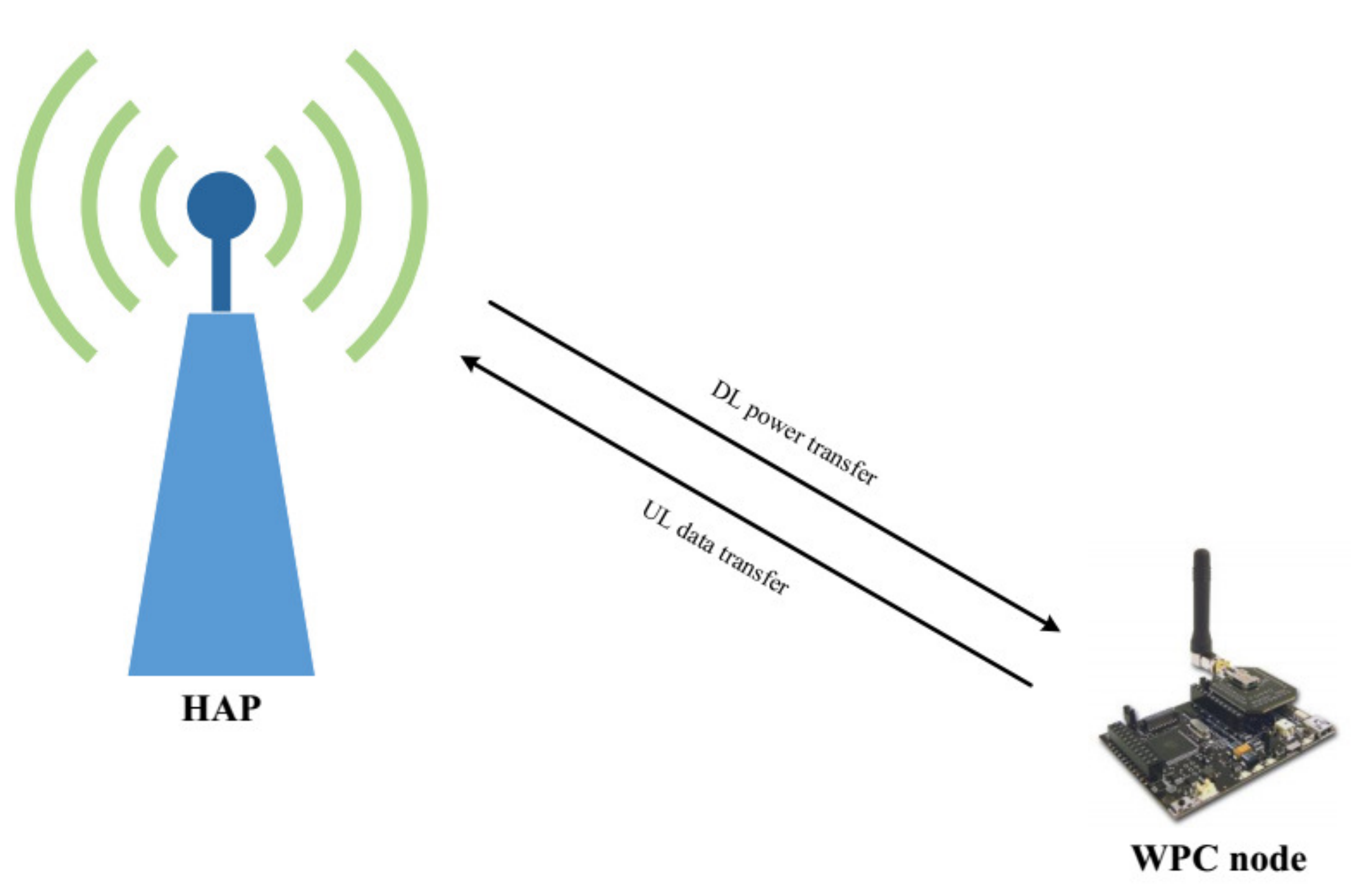}
	\end{center}
	\caption{System model.}
	\label{fig:1}
\end{figure}

We consider a point-to-point WPC system in \nakagami flat fading channel which include a HAP and a WPC node as shown in Fig. \ref{fig:1}. The HAP is equipped with a power amplifier with the average power of $P_h$. So, when the node has some new data to send, i.e. a frame of $R_0$ bits, the HAP starts transmitting power to the node for specified time duration $T_1$ in the DL phase. The HAP transmits its power as $\beta P_h$, where $\beta$ indicates the allocated power coefficient. Hence, in the ordinary way and without power allocation (problems $\mathcal{P}_1$ and $\mathcal{P}_3$) we assume $\beta=1$. However, when the optimal power allocation is assumed (Problems $\mathcal{P}_2$ and $\mathcal{P}_4$), $\beta$ can vary with the constant average and $\mathbb{E}\{\beta\}\leq 1$. At the end of DL time duration, the harvested energy at the WPC node is equal to
\begin{equation}\label{eq:node_energy}
E_n=\eta\beta P_h d^{-\alpha} |h_1|^2T_1
\end{equation}
where $0\leq\eta\leq 1$ is the energy harvesting efficiency, $d$ shows the distance between the HAP and node, $\alpha$ denotes the path-loss exponent, $h_1$ is the small scale DL channel coefficient and $\mathbb{E}\{|h_1|^2\}=1$. 

After the DL phase, the WPC node transmits $R_0$ bits to the HAP in the UL phase in the allocated time duration $T_2$. The node uses the harvested energy within $T_2$ amount of time, so, the transmitted power from the node is given by
\begin{equation}\label{eq:node_power}
P_n=\frac{E_n}{T_2}.
\end{equation}
In the time duration $T_2$, the WPC node can transmit $R_n$ bits such as
\begin{equation}\label{eq:node_rate_1}
R_n=BT_2\log_2\left(1+\frac{P_n d^{-\alpha}|h_2|^2}{\sigma_h^2}\right)
\end{equation}
where $B$ is the overall bandwidth, $h_2$ represents the small scale UL channel coefficient and $\sigma_h^2$ shows the noise power at the HAP respectively. For successful transmission at the UL, the channel capacity has to be greater than the available data so $R_n\geq R_0$. Without loss of generalities, we can assume slow fading in the following where $h_1=h_2=h_0$. Now, from \eqref{eq:node_energy}--\eqref{eq:node_rate_1}, we can summarize the instantaneous transmitted rate from the node to the HAP as
\begin{equation}\label{eq:node_rate_2}
R_n
=BT_2\log_2\left(1+\frac{\eta\beta P_h d^{-2\alpha}|h_0|^4T_1}{T_2\sigma_h^2}\right)
=BT_2\log_2\left(1+\beta a_0\h^2\frac{T_1}{T_2}\right)
\end{equation}
where $a_0=\eta P_h d^{-2\alpha}/\sigma_h^2$ is considered as the average signal to noise ratio (SNR) at the HAP and $\h=|h_0|^2$ represents the channel power gain. Since we assumed \nakagami flat fading channel and $\mathbb{E}\{\h\}=1$, we can write the probability density function (PDF) of $\h$ as \cite{Alouini}
\begin{equation}\label{eq:nakagami_m_pdf}
f_{\h}(x)=\frac{m^m x^{m-1}\e^{-m x}}{\Gamma(m)},~x\geq 0
\end{equation}
where $m$ is the \nakagami fading parameter and $\Gamma(.)$ shows the gamma function \cite{Ryzhik}.


\section{Transmission Delay Minimization}\label{sec:rtt_min}

As we explained before, data transmission with the low latency is required. So, our optimization problems are exactly considered this kind of delay $T_1+T_2$, which includes time duration of power transfer from the HAP to the node ($T_1$) and time duration of data transfer from the node to the HAP ($T_2$). In four following problems, we study transmission delay (TD) minimization with different assumptions respectively.
\subsection{Problem $\mathcal{P}_1$}\label{subsec:problem_p1}
In the first optimization problem, the HAP transmits constant power, so, the allocated power coefficient $\beta=1$. In addition we assume $T_1=T_2=T_0$. Then, the total TD is $T_1+T_2=2T_0$ and we can write the minimization problem as
\begin{equation}\label{eq:opt_p1}
T_0^*=\arg\mathop{\min}_{T_0}~2T_0
\end{equation}
\begin{equation}\label{eq:const_opt_p1}
\mathrm{s.t.}~~ R_n=BT_0\log_2\left(1+a_0\h^2\right)\geq R_0
\end{equation}
where \eqref{eq:opt_p1} and \eqref{eq:const_opt_p1} represent the cost function and the constraint of $\mathcal{P}_1$ and $T_0^*$ shows the optimal solution of the DL and UL time duration.

With regard equality in \eqref{eq:const_opt_p1}, the optimization problem $\mathcal{P}_1$ can be solved simply and $T_0^*$ is obtained as
\begin{equation}\label{eq:solution_opt_p1}
T_0^*=\frac{R_0}{B\log_2\left(1+a_0\h^2\right)}.
\end{equation}
Then, the average TD is calculated and given by 
\begin{equation}\label{eq:mean_rtt_opt_p1}
\overline{\textsf{TD}}=\int_0^{\infty}2T_0^*f_{\h}(x)dx=\int_0^{\infty}2\frac{R_0}{B\log_2\left(1+a_0x^2\right)}f_{\h}(x)dx
\end{equation}
where $f_{\h}(x)$ is written in \eqref{eq:nakagami_m_pdf} previously. Note that, the integral of \eqref{eq:mean_rtt_opt_p1} can be calculated numerically for $m>2$. For the proof, please refer to \ref{apndx:A}.

\subsection{Problem $\mathcal{P}_2$}\label{subsec:problem_p2}
In the second optimization problem when we use the optimal power allocation at the HAP, we can assume $T_1=T_2=T_0$ and the allocated power coefficient $\beta$ in \eqref{eq:node_rate_2} is $\beta\neq1$. However, in order to keep the average transmitted power from the HAP less or equal than $P_h$, we have to force $\mathbb{E}\{\beta\}\leq 1$. Now, we can write the minimization problem as
\begin{equation}\label{eq:opt_p2}
\left\{T_0^*,\beta^*\right\}=\arg\mathop{\min}_{T_0,\beta}~2T_0
\end{equation}
\begin{subequations}\label{eq:const_opt_p2}
	\begin{align}
	\label{eq:const_opt_p2_a}
	\mathrm{s.t.}~~ &R_n=BT_0\log_2\left(1+\beta a_0\h^2\right)\geq R_0
	\\                                                                      
	\label{eq:const_opt_p2_b}
	&\mathbb{E}\{\beta\}\leq 1
	\end{align}
\end{subequations}
where \eqref{eq:opt_p2} is the cost function and \eqref{eq:const_opt_p2_a}, \eqref{eq:const_opt_p2_b} represent the constraints of $\mathcal{P}_2$. 

After that, using constraint optimization method such as Lagrangian multiplier rule, the optimal solutions can be found as
\begin{equation}\label{eq:solution_beta_opt_p2}
\beta^*=\dfrac{\e^{2\mathcal{W}_0\left(\sqrt{\dfrac{2\ln2a_0\h^2R_0}{B\mu^*}}\Bigg/2\right)}-1}{a_0\h^2}
\end{equation}
and
\begin{equation}\label{eq:solution_T0_opt_p2}
T_0^*=\dfrac{R_0\ln2}{B\ln\left(1+\beta^*a_0\h^2\right)}=\dfrac{R_0\ln2}{2B\mathcal{W}_0\left(\sqrt{\dfrac{2\ln2a_0\h^2R_0}{B\mu^*}}\Bigg/2\right)}
\end{equation}
where $\mathcal{W}_0(.)$ indicates Lambert-W function \cite{lambertw}. For the details of derivation, please see \ref{apndx:B}. In addition, $\mu^*$ in \eqref{eq:solution_beta_opt_p2} and \eqref{eq:solution_T0_opt_p2} is a constant (Lagrange multiplier) and can be obtained from 
\begin{equation}\label{eq:unity_power_opt_p2}
\mathbb{E}\{\beta^*\}=\int_0^{\infty}\beta^*f_{\h}(x)dx=\int_0^{\infty}\dfrac{\e^{2\mathcal{W}_0\left(\sqrt{\dfrac{2\ln2a_0x^2R_0}{B\mu^*}}\Bigg/2\right)}-1}{a_0x^2}f_{\h}(x)dx=1.
\end{equation}
In a similar way, the average TD is given by
\begin{equation}\label{eq:mean_rtt_opt_p2}
\overline{\textsf{TD}}=\int_0^{\infty}2T_0^*f_{\h}(x)dx=\int_0^{\infty}\dfrac{2R_0\ln2}{2B\mathcal{W}_0\left(\sqrt{\dfrac{2\ln2a_0x^2R_0}{B\mu^*}}\Bigg/2\right)}f_{\h}(x)dx.
\end{equation}
Note that, the integrals of \eqref{eq:unity_power_opt_p2} and  \eqref{eq:mean_rtt_opt_p2} can be calculated numerically when $m>2$ (see \ref{apndx:C}).

\subsection{Problem $\mathcal{P}_3$}\label{subsec:problem_p3}
Here we assume a general case with $T_1\neq T_2$ but without the optimal power allocation at the HAP. So, $\beta=1$ in \eqref{eq:node_rate_2} and the minimization problem is formulated as
\begin{equation}\label{eq:opt_p3}
\left\{T_1^*,T_2^*\right\}=\arg\mathop{\min}_{T_1,T_2}~T_1+T_2
\end{equation}
\begin{equation}\label{eq:const_opt_p3}
\mathrm{s.t.}~~ R_n=BT_2\log_2\left(1+a_0\h^2\dfrac{T_1}{T_2}\right)\geq R_0
\end{equation}
where \eqref{eq:opt_p3} indicates the cost function and \eqref{eq:const_opt_p3} represents the constraint of $\mathcal{P}_3$. Solving the constraint optimization problem, we can find
\begin{equation}\label{eq:solution_T1_opt_p3}
T_1^*=\dfrac{R_0\ln2}{B\left(1+\mathcal{W}_0\left(\dfrac{a_0\h^2-1}{\e}\right)\right)}\dfrac{\e^{1+\mathcal{W}_0\left(\dfrac{a_0\h^2-1}{\e}\right)}-1}{a_0\h^2}
\end{equation}
\begin{equation}\label{eq:solution_T2_opt_p3}
T_2^*=\dfrac{R_0\ln2}{B\left(1+\mathcal{W}_0\left(\dfrac{a_0\h^2-1}{\e}\right)\right)}
\end{equation}
and the average TD is calculated according to
\begin{equation}\label{eq:mean_rtt_opt_p3}
\overline{\textsf{TD}}=\int_0^{\infty}\left(T_1^*+T_2^*\right)f_{\h}(x)dx.
\end{equation}
Note that, the integral of \eqref{eq:mean_rtt_opt_p3} is convergent for $m>2$. The details of this problem solving are very similar to \ref{apndx:B} which is omitted here. In addition, the convergence test of \eqref{eq:mean_rtt_opt_p3} for $m>2$ can be obtained similar to \ref{apndx:A} or \ref{apndx:C}. 

\subsection{Problem $\mathcal{P}_4$}\label{subsec:problem_p4}
Problem $\mathcal{P}_4$ is the most general case where we assume $T_1\neq T_2$ and the optimal power allocation coefficient $\beta$. So, the minimization problem is formulated as
\begin{equation}\label{eq:opt_p4}
\left\{T_1^*,T_2^*,\beta^*\right\}=\arg\mathop{\min}_{T_1,T_2,\beta}~T_1+T_2
\end{equation}
\begin{subequations}\label{eq:const_opt_p4}
	\begin{align}
	\label{eq:const_opt_p4_a}
	\mathrm{s.t.}~~ &R_n=BT_2\log_2\left(1+\beta a_0\h^2\dfrac{T_1}{T_2}\right)\geq R_0
	\\                                                                      
	\label{eq:const_opt_p4_b}
	&\mathbb{E}\{\beta\}\leq 1
	\end{align}
\end{subequations}
where \eqref{eq:opt_p4} is the cost function and \eqref{eq:const_opt_p4_a}, \eqref{eq:const_opt_p4_b} represent the constraints of $\mathcal{P}_4$ respectively. Similar to the previous cases, using the Lagrangian multiplier rule, the optimal solution is calculated numerically. Here, $\beta^*$ is obtained from
\begin{equation}\label{eq:solution_beta_opt_p4}
\dfrac{B\mu^*a_0\h^2}{R_0\ln2}\beta^{*2}\left(1+\mathcal{W}_0\left(\dfrac{\beta^*a_0\h^2-1}{\e}\right)\right)=\e^{1+\mathcal{W}_0\left(\dfrac{\beta^*a_0\h^2-1}{\e}\right)}-1
\end{equation}
and then, 
\begin{equation}\label{eq:solution_T1_opt_p4}
T_1^*=\beta^*\mu^*
\end{equation}
\begin{equation}\label{eq:solution_T2_opt_p4}
T_2^*=\dfrac{\beta^{*2}\mu^*a_0\h^2}{\e^{1+\mathcal{W}_0\left(\dfrac{\beta^*a_0\h^2-1}{\e}\right)}-1}
\end{equation}
where $\mu^*$ is a constant (Lagrange multiplier) and obtained from $\mathbb{E}\{\beta^*\}=1$. For the details, please see \ref{apndx:D}. Since \eqref{eq:solution_beta_opt_p4} is highly nonlinear, the computation of $\mu^*$ is performed iteratively. This means that, we first set $\mu^*$ and find $\beta^*$ from \eqref{eq:solution_beta_opt_p4} when $\h$ changes over $(0,\infty)$. Then, $\mathbb{E}\{\beta^*\}$ is calculated and compared with the optimal value $\mathbb{E}\{\beta^*\}=1$ and then, $\mu^*$ is updated as well. After that, we find $\beta^*$ from \eqref{eq:solution_beta_opt_p4} again and this process continuous until $\mathbb{E}\{\beta^*\}\approx 1$. Accordingly, we can observe that calculation of $\mu^*$ and $\beta^*$ from \eqref{eq:solution_beta_opt_p4} is so complex and time consuming. Therefore, we suggest an approximation to reduce this complexity as follows.

Again, we consider \eqref{eq:solution_beta_opt_p4} and replace $\beta^*$ in the argument of Lambert-W function with its average value $\mathbb{E}\{\beta^*\}=1$ and simplify \eqref{eq:solution_beta_opt_p4} as
\begin{equation}\label{eq:approx_1_beta_opt_p4}
\dfrac{B\mu^*a_0\h^2}{R_0\ln2}\beta^{*2}\left(1+\mathcal{W}_0\left(\dfrac{1\times a_0\h^2-1}{\e}\right)\right)=\e^{1+\mathcal{W}_0\left(\dfrac{1\times a_0\h^2-1}{\e}\right)}-1.
\end{equation}
Now, we have a closed-form solution for $\beta^*$ given by
\begin{equation}\label{eq:approx_2_beta_opt_p4}
\beta^*=\sqrt{\dfrac{R_0\ln2}{\mu^*B}}\sqrt{\dfrac{\e^{1+\mathcal{W}_0\left(\dfrac{ a_0\h^2-1}{\e}\right)}-1}{a_0\h^2\left(1+\mathcal{W}_0\left(\dfrac{a_0\h^2-1}{\e}\right)\right)}}
\end{equation}
and accordingly
\begin{equation}\label{eq:approx_mu_opt_p4}
\mu^*=\dfrac{R_0\ln2}{B}\mathbb{E}\left\{\sqrt{\dfrac{\e^{1+\mathcal{W}_0\left(\dfrac{ a_0\h^2-1}{\e}\right)}-1}{a_0\h^2\left(1+\mathcal{W}_0\left(\dfrac{a_0\h^2-1}{\e}\right)\right)}}\right\}^2.
\end{equation}
Then, $T_1^*$ and $T_2^*$ are extracted from \eqref{eq:solution_T1_opt_p4} and \eqref{eq:solution_T2_opt_p4} respectively and the average TD is calculated by
\begin{equation}\label{eq:mean_rtt_opt_p4}
\overline{\textsf{TD}}=\int_0^{\infty}\left(T_1^*+T_2^*\right)f_{\h}(x)dx.
\end{equation}
Note that, we can show \eqref{eq:approx_mu_opt_p4} and \eqref{eq:mean_rtt_opt_p4} are convergent when we choose $m>2$. The proof is similar to \ref{apndx:A} and \ref{apndx:C} which is omitted for briefing.

Finally, after extracting the optimal solution of problems $\mathcal{P}_1$ to $\mathcal{P}_4$, we observe that the solution is not employable in the Rayleigh channel (Rayleigh channel is a subset of \nakagami when $m=1$). Therefore, some practical technique such as truncated power allocation \cite{Lari2} is recommended for the Rayleigh channel to resolve this issue. However, this issue is considered in a different paper in the future.


\section{Extension to The Multiuser Scenario}\label{sec:extension_multiuser}
In this section we attempt to extend the previous work to the multiuser case. So, here we assume one HAP and $K$ nodes. At the DL, the HAP transmits power to the nodes in time duration equals to $T_1$. $K$ nodes can harvest energy during this time and they become ready to transmit their information to the HAP in the UL. Without loss of generality, we assume all $K$ nodes have $R_0$~bits to transmit at the UL and transmit these information one by one in a TDMA slots. $T_{2,k}$, $k=1,2,...,K$ denotes a dedicated time slot to the $k$-th node in the UL. Similar to the single user case, $a_k=\eta P_h d_k^{-2\alpha}/\sigma_h^2$ represents the average SNR related to the $k$-th node at the HAP, $d_k$ indicates the distance between the $k$-th node and HAP and $\h_k$ shows the channel power gain between the the $k$-th node and HAP as well. The other parameters are the same as the single user case. In the multiuser case, one complete transmission in the DL and UL carry outs in time duration of $T_1+T_{2,1}+T_{2,2}+...+T_{2,K}$. Therefore, minimization of $T_1+T_{2,1}+T_{2,2}+...+T_{2,K}$ is our concern here.


\subsection{Problem $\mathcal{P}_5$}\label{subsec:problem_p5}
Here we assume that $T_{2,1}+T_{2,2}+...+T_{2,K}= T_1$. By this assumption, when the number of nodes $K$ changes, the total UL time remains equal to the DL time $T_1$ and this is more realistic in a TDMA based system. For more simplicity, power allocation is not applied and therefore, in this problem we have $\beta=1$. So, the total TD is $T_1+T_{2,1}+T_{2,2}+...+T_{2,K}$ and we can write the minimization problem as
\begin{equation}\label{eq:opt_p5}
\{T_1^*,T_{2,1}^*,T_{2,2}^*,...T_{2,K}^*\}=\arg\mathop{\min}_{T_1,T_{2,1},T_{2,2},...+T_{2,K}}~T_1+T_{2,1}+T_{2,2}+...+T_{2,K}
\end{equation}
\begin{subequations}\label{eq:const_opt_p5}
	\begin{align}
	\label{eq:const_opt_p5_a}
	\mathrm{s.t.}~~ &R_{n,1}=BT_{2,1}\log_2\left(1+a_1\h_1^2\frac{T_1}{T_{2,1}}\right)\geq R_0
	\\
	\label{eq:const_opt_p5_b}
	&R_{n,2}=BT_{2,2}\log_2\left(1+a_2\h_2^2\frac{T_1}{T_{2,2}}\right)\geq R_0                                                                     
	\\
	&...
	\\
	\label{eq:const_opt_p5_K}
	&R_{n,K}=BT_{2,K}\log_2\left(1+a_K\h_K^2\frac{T_1}{T_{2,K}}\right)\geq R_0
	\\
	\label{eq:const_opt_p5_L}
	&T_{2,1}+T_{2,2}+...+T_{2,K}= T_1
	\end{align}
\end{subequations}
where \eqref{eq:opt_p5} is the cost function and \eqref{eq:const_opt_p5_a}-\eqref{eq:const_opt_p5_L} represent the constraints of $\mathcal{P}_5$. 

Problem $\mathcal{P}_5$ can be solved using Lagrange multiplier rule. However, obtaining an optimal closed-form solution for the $K+1$ unknowns $T_1,T_{2,1}+T_{2,2}+...+T_{2,K}$ is not possible and we can find them numerically from $K+1$ nonlinear equations
\begin{subequations}\label{eq:solution_opt_p5}
	\begin{align}
	\label{eq:solution_T2_opt_p5_a}
	&BT_{2,1}^*\log_2\left(1+a_1\h_1^2\frac{T_1^*}{T_{2,1}^*}\right)=R_0
	\\
	\label{eq:solution_T2_opt_p5_b}
	&BT_{2,2}^*\log_2\left(1+a_2\h_2^2\frac{T_1^*}{T_{2,2}^*}\right)=R_0
	\\
	&...
	\\
	\label{eq:solution_T2_opt_p5_K}
	&BT_{2,K}^*\log_2\left(1+a_K\h_K^2\frac{T_1^*}{T_{2,K}^*}\right)=R_0
	\\
	\label{eq:solution_T1_opt_p5}
	&T_{2,1}^*+T_{2,2}^*+...+T_{2,K}^*=T_1^*
	\end{align}
\end{subequations}
where $T_1^*$,$T_{2,1}^*$,$T_{2,2}^*$,$...$,$T_{2,K}^*$ denote the optimal values for the DL and UL time durations. The details for derivation of \eqref{eq:solution_T2_opt_p5_a}-\eqref{eq:solution_T1_opt_p5} is very similar to \ref{apndx:B} and \ref{apndx:D} and it is not repeated again here.
\subsection{Problem $\mathcal{P}_6$}\label{subsec:problem_p6}
In the problem $\mathcal{P}_6$, we extend $\mathcal{P}_5$ and add power allocation at the HAP for more improvement. Therefore, $\beta\neq 1$ and for keeping the average transmitted power from the HAP less or equal to $P_h$, we have $\mathbb{E}\{\beta\}\leq 1$. Again we assume $T_{2,1}+T_{2,2}+...+T_{2,K}=T_1$ in the UL. Then, the minimization problem is written as
\begin{eqnarray}\label{eq:opt_p6}
\{T_1^*,T_{2,1}^*,T_{2,2}^*,...T_{2,K}^*,\beta^*\}=\arg\mathop{\min}_{T_1,T_{2,1},T_{2,2},...+T_{2,K},\beta}~
T_1+T_{2,1}+T_{2,2}+...+T_{2,K}
\end{eqnarray}
\begin{subequations}\label{eq:const_opt_p6}
	\begin{align}
	\label{eq:const_opt_p6_a}
	\mathrm{s.t.}~~ &R_{n,1}=BT_{2,1}\log_2\left(1+a_1\h_1^2\beta\frac{T_1}{T_{2,1}}\right)\geq R_0
	\\
	\label{eq:const_opt_p6_b}
	&R_{n,2}=BT_{2,2}\log_2\left(1+a_2\h_2^2\beta\frac{T_1}{T_{2,2}}\right)\geq R_0                                                                                                                                                                                                                                                                                                                                                                                                                                                                                                                                                                                                                                                                                                                                                                                                                                                                                                                                                                                                                                                                                                                                                                                                                                                                                                                                                                                                                                                                                                                                                                                                                                                                                                                                                                                                                                                                                                                                                                                                                                                                                                                                                                                                                                                                                                                                                                                                                                                                                                                                                                                                                                                                                                                                                                                                                                                                                                                                                                                                                                                                                                                                                                                                                                                                                                                                                                                                                                                                                                                                                                                                                                                                                                                                           
	\\
	&...
	\\
	\label{eq:const_opt_p6_K}
	&R_{n,K}=BT_{2,K}\log_2\left(1+a_K\h_K^2\beta\frac{T_1}{T_{2,K}}\right)\geq R_0
	\\
	\label{eq:const_opt_p6_L}
	&T_{2,1}+T_{2,2}+...+T_{2,K}=T_1
	\\
	\label{eq:const_opt_p6_M}
	&\mathbb{E}\{\beta\}\leq 1
	\end{align}
\end{subequations}
where \eqref{eq:opt_p6} is the cost function and \eqref{eq:const_opt_p6_a}-\eqref{eq:const_opt_p6_M} show the constraints. Using Lagrange multiplier method, the optimal values are derived numerically from $2K+4$ equations as follows:
\begin{subequations}\label{eq:solution_opt_p6}
	\begin{align}
	\label{eq:solution_opt_p6_1}
	&\mu^*+\lambda_1^*\frac{\frac{B}{\ln2}a_1\h_1^2\beta^*}{1+a_1\h_1^2\beta^*\frac{T_1^*}{T_{2,1}^*}}+...+\lambda_K^*\frac{\frac{B}{\ln2}a_K\h_K^2\beta^*}{1+a_K\h_K^2\beta^*\frac{T_1^*}{T_{2,K}^*}}=1
	\\
	\label{eq:solution_opt_p6_2}
	&-\mu^*+\lambda_1^*\left(B\log_2\left(1+a_1\h_1^2\beta^*\frac{T_1^*}{T_{2,1}^*}\right)-\frac{Ba_1\h_1^2\beta^*\frac{T_1^*}{T_{2,1}^*}}{\ln2\left(1+a_1\h_1^2\beta^*\frac{T_1^*}{T_{2,1}^*}\right)}\right)=1
	\\
	&...
	\\
	\label{eq:solution_opt_p6_3}
	&-\mu^*+\lambda_K^*\left(B\log_2\left(1+a_K\h_K^2\beta^*\frac{T_1^*}{T_{2,K}^*}\right)-\frac{Ba_K\h_K^2\beta^*\frac{T_1^*}{T_{2,K}^*}}{\ln2\left(1+a_K\h_K^2\beta^*\frac{T_1^*}{T_{2,K}^*}\right)}\right)=1
	\\
	\label{eq:solution_opt_p6_4}
	&\lambda_1^*\frac{Ba_1\h_1^2T_1^*}{\ln2\left(1+a_1\h_1^2\beta^*\frac{T_1^*}{T_{2,1}^*}\right)}+...+\lambda_K^*\frac{Ba_K\h_K^2T_1^*}{\ln2\left(1+a_K\h_K^2\beta^*\frac{T_1^*}{T_{2,K}^*}\right)}-\theta^*=0
	\\
	\label{eq:solution_opt_p6_5}
	&BT_{2,1}^*\log_2\left(1+a_1\h_1^2\beta^*\frac{T_1^*}{T_{2,1}^*}\right)=R_0
	\\
	...
	\\
	\label{eq:solution_opt_p6_6}
	&BT_{2,K}^*\log_2\left(1+a_K\h_K^2\beta^*\frac{T_1^*}{T_{2,K}^*}\right)=R_0
	\\
	\label{eq:solution_opt_p6_7}
	&\mu^*\left(T_1^*+T_{2,1}^*+...+T_{2,K}^*\right)=0
	\\
	\label{eq:solution_opt_p6_8}
	&\mathbb{E}\{\beta^*\}=1
	\end{align}
\end{subequations}
where $T_1^*$,$T_{2,1}^*$,$T_{2,2}^*$,...,$T_{2,K}^*$ denote the optimal values for the DL and UL time duration and $\beta^*$ represents the optimal allocated power coefficient respectively. In addition, $\lambda_1^*$,$\lambda_2^*$,...,$\lambda_K^*$ represent the optimal values for Lagrange multiplier corresponding to the constraints \eqref{eq:const_opt_p6_a}-\eqref{eq:const_opt_p6_K}, $\mu^*$ shows the optimal value for Lagrange multiplier corresponding to the constraint \eqref{eq:const_opt_p6_L} and $\theta^*$ shows the optimal value for Lagrange multiplier corresponding to the constraint \eqref{eq:const_opt_p6_M}. The details for derivation of \eqref{eq:solution_opt_p6_1}-\eqref{eq:solution_opt_p6_8} are very similar to \ref{apndx:B} and \ref{apndx:D} and therefore it is not repeated again.

The system of $2K+4$ equations with $2K+4$ unknowns in \eqref{eq:solution_opt_p6_1}-\eqref{eq:solution_opt_p6_8} can be solved iteratively. The steps are shown in Table \ref{alg:solution_p6}. First we initialize $\theta^*>0$ and we solve \eqref{eq:solution_opt_p6_1}-\eqref{eq:solution_opt_p6_7} for different values of $\{\h_1,\h_2,...,\h_K\}$ change over $(0,\infty)$. Then, the average $\mathbb{E}\{\beta^*\}$ is calculated as Monte-Carlo average and compared with the optimal value $\mathbb{E}\{\beta^*\}=1$.
If this average is almost one, \eqref{eq:solution_opt_p6_8} is satisfied and $T_1^*$,$T_{2,1}^*$,...,$T_{2,K}^*$ and $\beta^*$ show the optimal values. Otherwise, $\theta^*$ is updated with sub-gradient method \cite{Sub-Gradient} in \eqref{eq:sub_gradient} and the iteration continues until \eqref{eq:solution_opt_p6_8} is satisfied. Note that, $\theta^{*(i)}$ in \eqref{eq:sub_gradient} shows the value at the $i$-th iteration, $\zeta>0$ is a positive gradient search step size and $[x]^+$ denotes $\max(0,x)$.
\begin{equation}\label{eq:sub_gradient}
\theta^{*(i+1)}=\left[\theta^{*(i)}-\zeta\left(\mathbb{E}\{\beta^*\}-1\right)\right]^+
\end{equation}
\begin{algorithm}\label{alg:solution_p6}
	\DontPrintSemicolon
	\SetKwInOut{Input}{input}
	\Input{$a_1>0$, $a_2>0$,...,$a_K>0$.}
	\Input{$\theta^*>0$ randomly.}
	\Input{$\epsilon$ as a very small number such as $10^{-6}$.}
	\BlankLine
	\While{$error>\epsilon$}{
		\For{$i=1:1:10000$}{
			Generate $\{\h_1,\h_2,...,\h_K\}$ over $(0,\infty)$ according to their distribution in \eqref{eq:nakagami_m_pdf}.\;
			Solve system of $2K+3$ nonlinear equations \eqref{eq:solution_opt_p6_1}-\eqref{eq:solution_opt_p6_7} numerically and find $2K+3$ unknowns $T_1^*$, $T_{2,1}^*$,...,$T_{2,K}^*$, $\lambda_1^*$,...,$\lambda_K^*$, $\mu^*$ and $\beta^*$.\;
			$b_i\leftarrow \beta^*$.\;
		}
		Calculate $\mathbb{E}\{\beta^*\}$ numerically as $\mathbb{E}\{\beta^*\}=\frac{b_1+b_2+...+b_{10000}}{10000}$.\;
		$error\leftarrow|\mathbb{E}\{\beta^*\}-1|$.	\;
		Update $\theta^*$ by \eqref{eq:sub_gradient}.\;
	}
	$T_1^*$, $T_{2,1}^*$,...,$T_{2,K}^*$ and $\beta^*$ are valid and represent the optimal solutions.\;
	\BlankLine
	\caption{Iterative solution for the problem $\mathcal{P}_6$}
\end{algorithm}

\section{Simulation Results}\label{sec:simulation_results}
In this section, the obtained results from four optimization problems $\mathcal{P}_1$ to $\mathcal{P}_4$ and also two optimization problems $\mathcal{P}_5$ to $\mathcal{P}_6$ are considered and compared to illustrate the benefits of optimal time and power allocation. Here we assume $B=100$KHz, $m=4$ and other parameters and variables will be specified if necessary. We have used MATLAB software package for the numerical results and for satisfactory results, Mont-Carlo simulation is repeated 100,000 times in each step.

\begin{figure}
	\begin{center}
		\includegraphics[draft=false,width=0.95\linewidth]{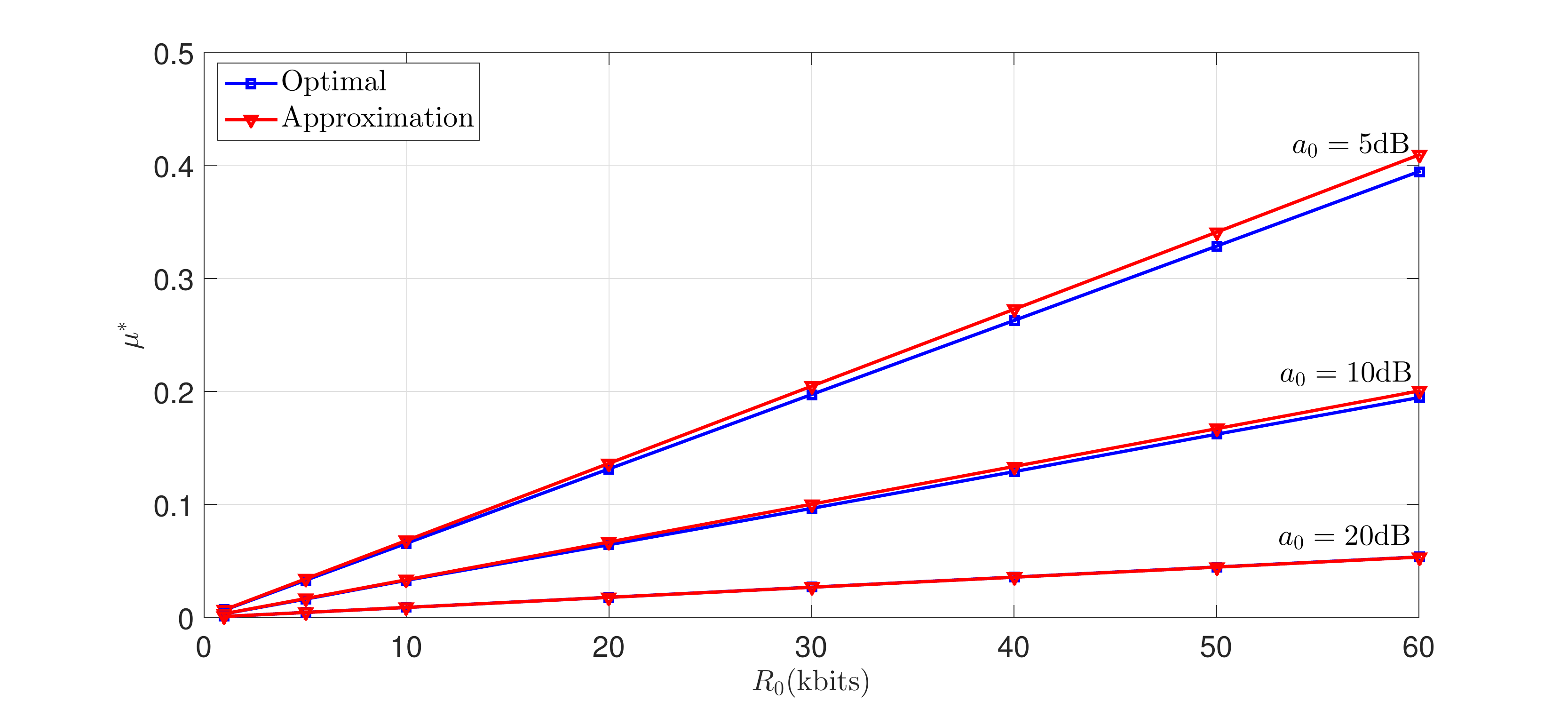}
	\end{center}
	\caption{The optimal and approximated values of $\mu^*$ in the problem $\mathcal{P}_4$.}
	\label{fig:2}
\end{figure}

First, we examine the accuracy of proposed approximation in the forth problem $\mathcal{P}_4$ in Fig. \ref{fig:2}. We plot the achieved values for $\mu^*$ from \eqref{eq:solution_beta_opt_p4} and \eqref{eq:approx_mu_opt_p4} simultaneously versus $R_0$ and for different values of the average SNR $a_0$. We can see an acceptable agreement between the optimal and approximated solutions, especially at high SNR and low values of $R_0$. Henceforth, due to the simplicity of \eqref{eq:approx_mu_opt_p4}, $\mu^*$ in problem $\mathcal{P}_4$ can be calculated by this equation.

\begin{figure}
	\begin{center}
		\includegraphics[draft=false,width=0.95\linewidth]{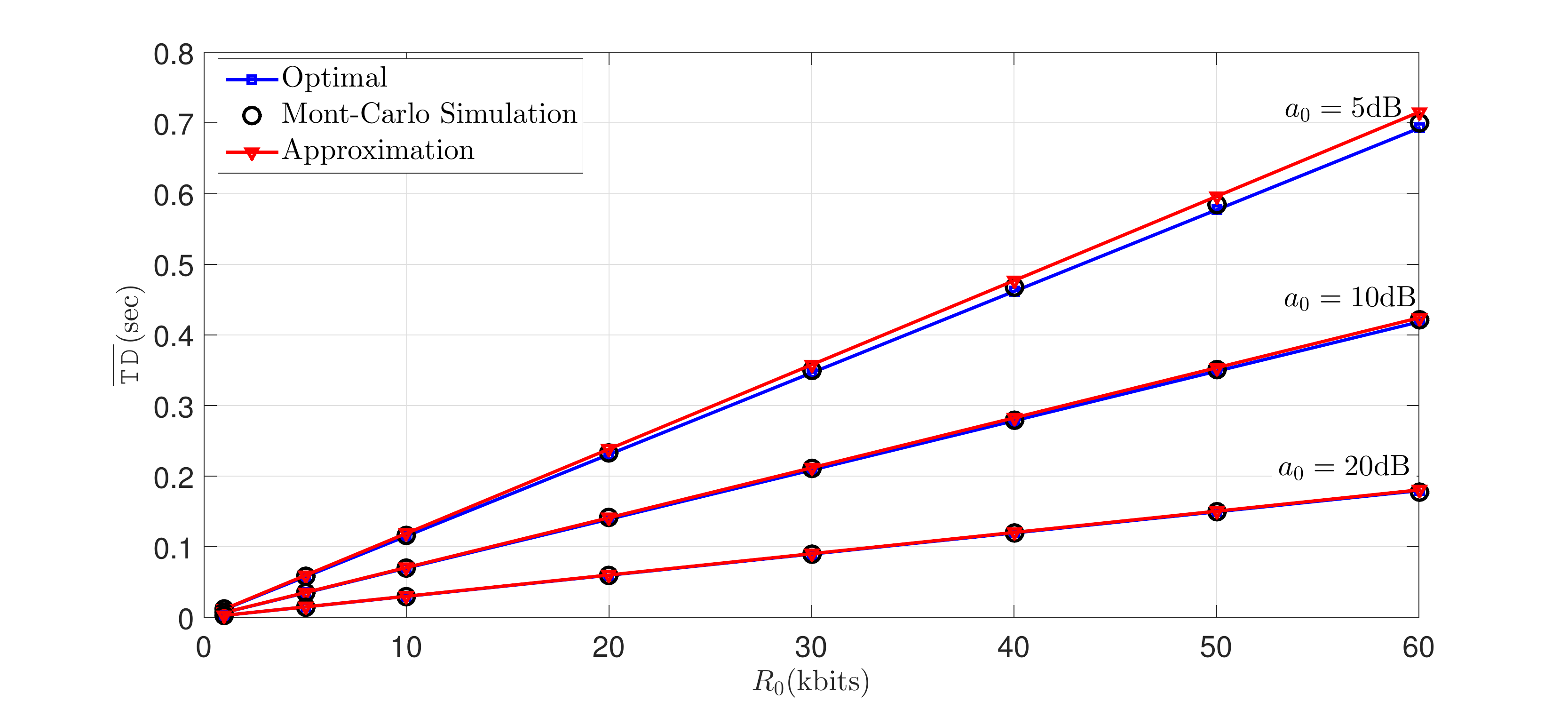}
	\end{center}
	\caption{Comparison of the optimal, approximated and Mont-Carlo simulation for $\overline{\textsf{TD}}$ in the problem $\mathcal{P}_4$.}
	\label{fig:3}
\end{figure}

In addition, Fig. \ref{fig:3} shows $\overline{\textsf{TD}}$ with the optimal and approximated solutions in problem $\mathcal{P}_4$ against $R_0$. For a better comparison, Mont-Carlo simulation is also added in this figure which has a great fit with the optimal solution. Again, an acceptable agreement between the optimal and approximated solutions can be observed. Furthermore, as we expected, $\overline{\textsf{TD}}$ decreases by increasing the average SNR $a_0$.

\begin{figure}
	\centering
	\subfloat[$a_0=5\textrm{dB}$]{\includegraphics[width=0.85\linewidth]{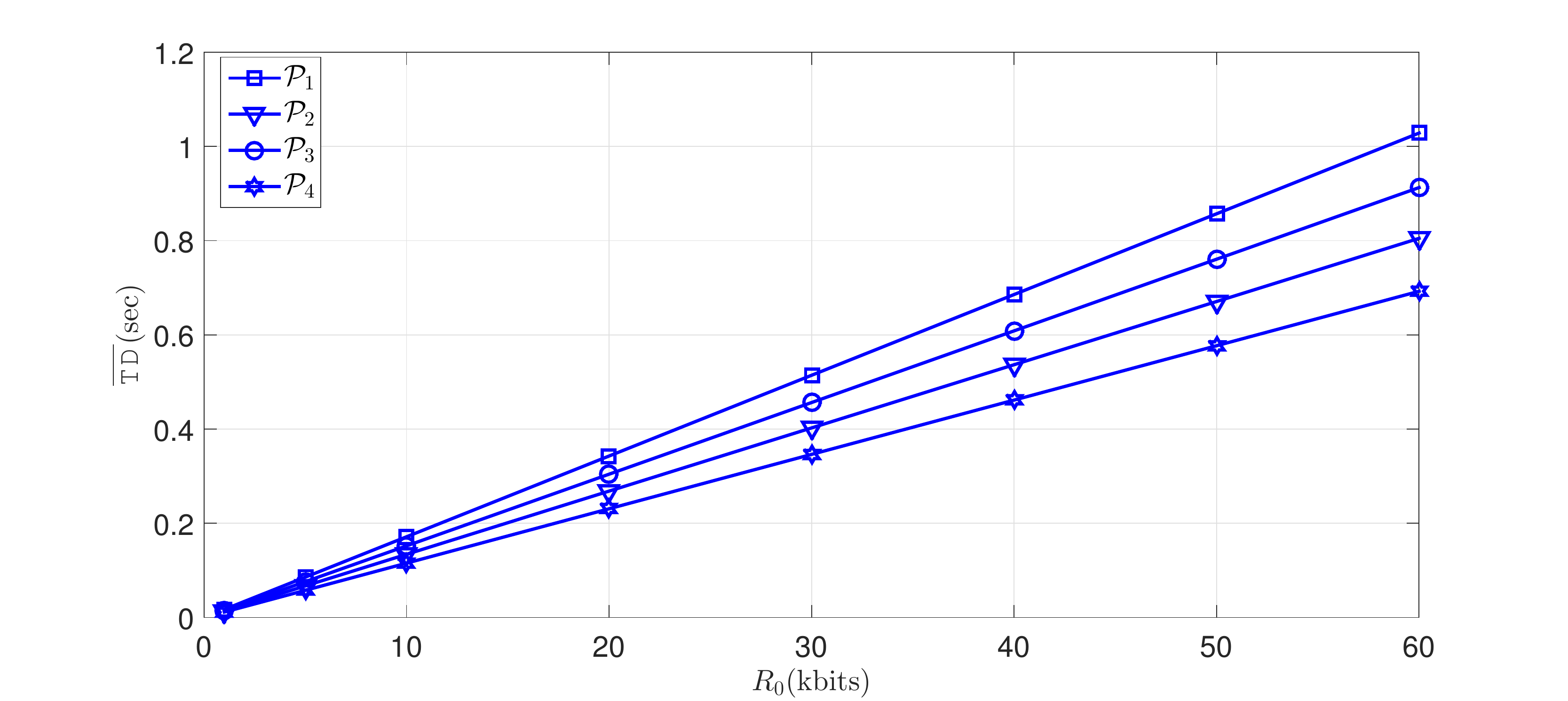}
		\label{fig:4a}}
	\hfil
	\subfloat[$a_0=10\textrm{dB}$]{\includegraphics[width=0.85\linewidth]{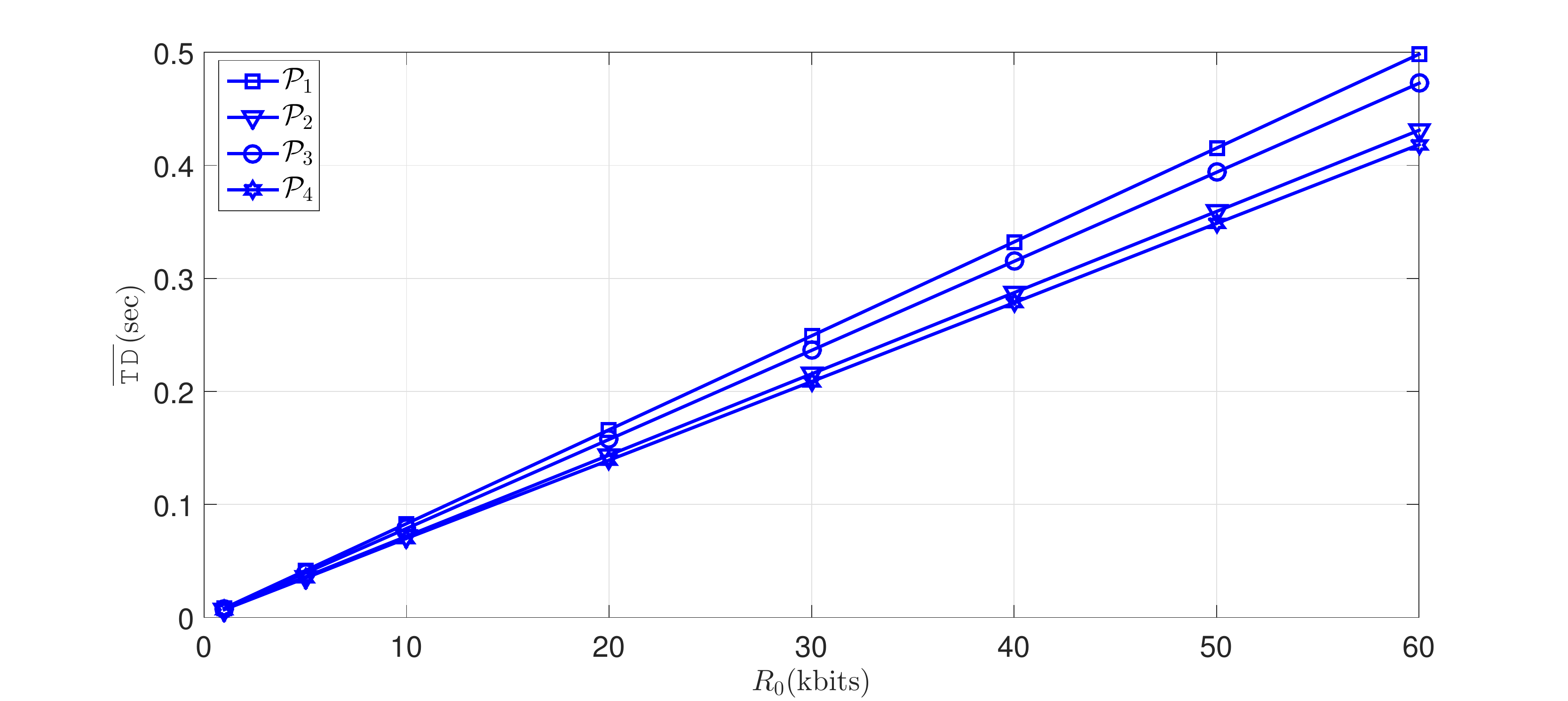}
		\label{fig:4b}}
	\hfil
	\subfloat[$a_0=20\textrm{dB}$]{\includegraphics[width=0.85\linewidth]{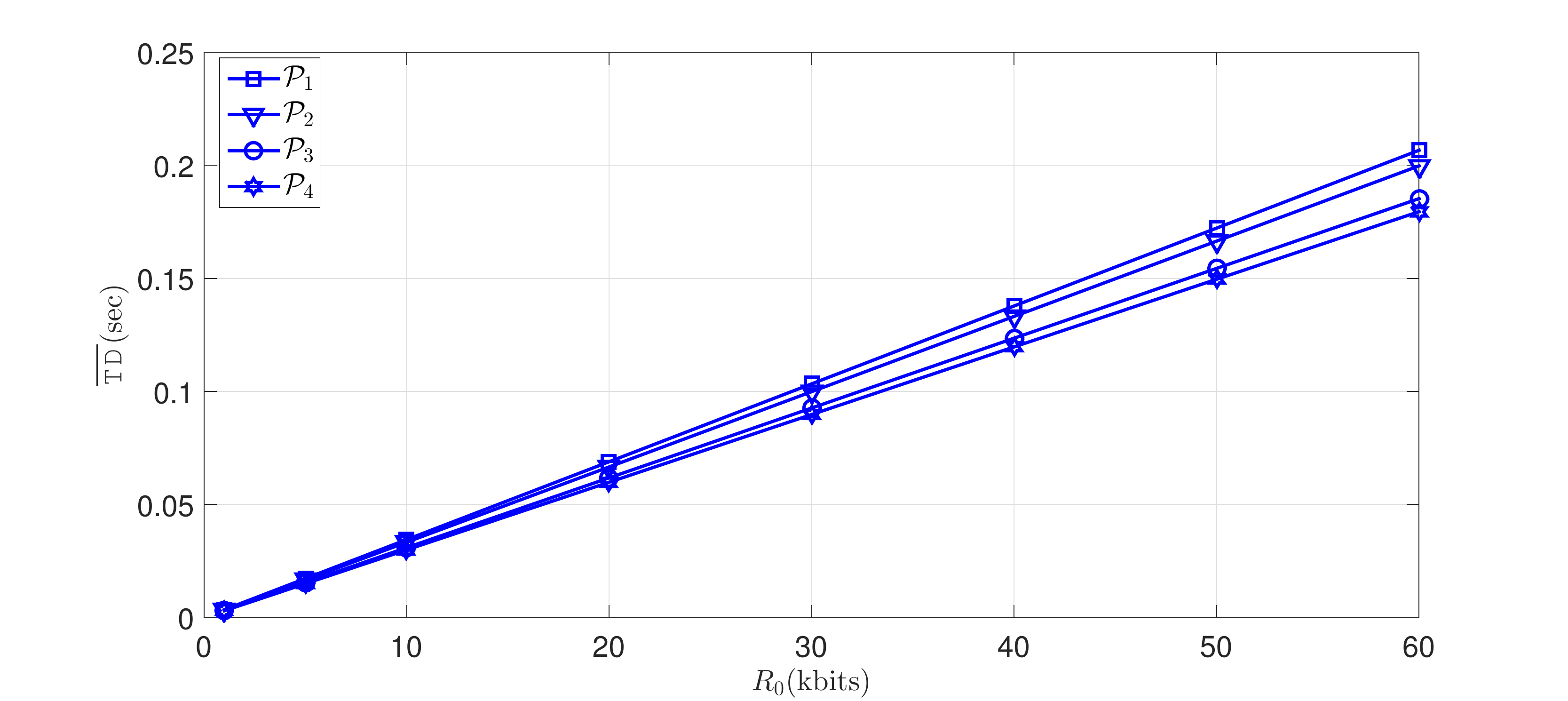}
		\label{fig:4c}}
	\caption{$\overline{\textsf{TD}}$ in different problems $\mathcal{P}_1$ to $\mathcal{P}_4$.}
	\label{fig:4}
\end{figure}

In Fig. \ref{fig:4} which can be considered as the main figure of this paper, the results of four optimization problems $\mathcal{P}_1$ to $\mathcal{P}_4$ are compared visually. In this figure, $\overline{\textsf{TD}}$ has been plotted versus $R_0$ whereas the average SNR $a_0$ takes different values (i.e. $a_0=5\textrm{dB}$ in Fig \ref{fig:4a}, $a_0=10\textrm{dB}$ in Fig \ref{fig:4b} and $a_0=20\textrm{dB}$ in Fig \ref{fig:4c}). As we expected, by increasing the average SNR $a_0$ in a given value of $R_0$, $\overline{\textsf{TD}}$ decreases as well. Then, comparing the solutions of four optimization problems reveals that the lowest value of $\overline{\textsf{TD}}$ is achieved from the results of problem $\mathcal{P}_4$. As the same way, the highest value of $\overline{\textsf{TD}}$ is obtained from the results of problem $\mathcal{P}_1$. Therefore, simultaneous time and power allocation similar to the problem $\mathcal{P}_4$ can reduce the total transmission delay significantly. For instance, when a WPC node has $R_0=50\textrm{kbits}$ data, Fig. \ref{fig:4a} shows that transmission with the optimal time and power allocation according to the problem $\mathcal{P}_4$ can reduce the delay by almost $34\%$ when compare to the problem $\mathcal{P}_1$. Consequently, applying the optimal time and power allocation is highly recommended especially for low and moderate SNRs. In addition, we can compare the solutions of $\mathcal{P}_2$ and $\mathcal{P}_3$ respectively. In the problem $\mathcal{P}_2$ we assumed an equal time duration at the DL and UL channels along with the optimal power allocation at the HAP. But in the problem $\mathcal{P}_3$, an unequal DL and UL time duration without the optimal power allocation are assumed. We can see that the obtained results from $\mathcal{P}_2$ are better than the results of $\mathcal{P}_3$. Consequently, we can conclude that the optimal power allocation has more influence on the delay rather than the optimal time allocation.

One other important observation in Fig. \ref{fig:4} is that when $a_0$ is increased, the results of problems $\mathcal{P}_1$ and $\mathcal{P}_2$ and also problems $\mathcal{P}_3$ and $\mathcal{P}_4$ tend to each other respectively. This is clearly visible in Fig. \ref{fig:4c}. Therefore, we can conclude that at high SNRs, the optimal power allocation does not have a favourable effect and against, the optimal time allocation has a better impact on the results as well. 
\begin{figure}
	\begin{center}
		\includegraphics[draft=false,width=0.95\linewidth]{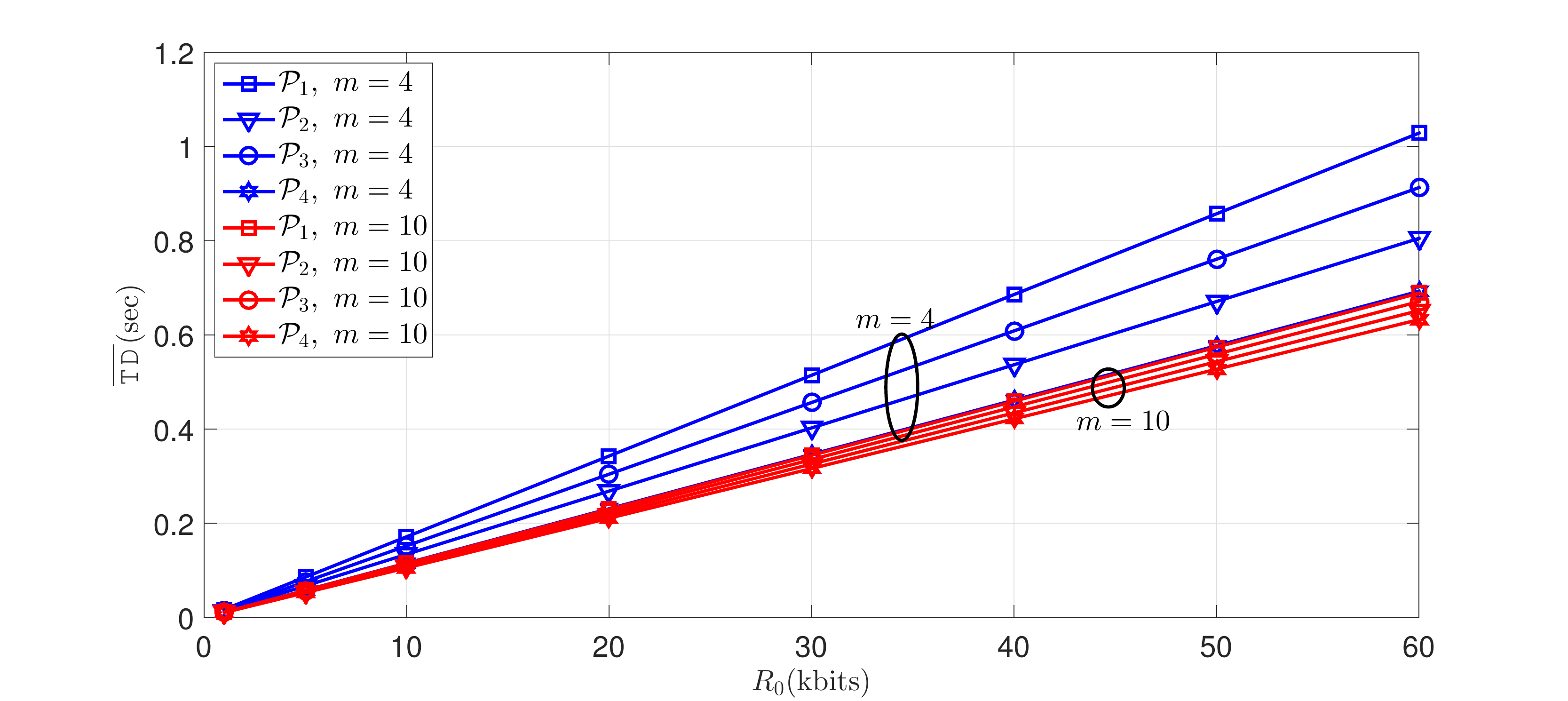}
	\end{center}
	\caption{Comparison of $\overline{\textsf{TD}}$ for different values of $m$.}
	\label{fig:5}
\end{figure}

At last, we plot $\overline{\textsf{TD}}$ for the problems $\mathcal{P}_1$ to $\mathcal{P}_4$ versus $R_0$ for $m=4$ and $m=10$ and $a_0=5$dB in Fig. \ref{fig:5}. The delay decreases when the fading parameter $m$ increases. As we can see, the optimal time and power allocation at high value of $m$, does not have a considerable impact on the performance. Therefore, using the optimal solution at lower value of $m$ is more required.

\begin{figure}
	\begin{center}
		\includegraphics[draft=false,width=0.95\linewidth]{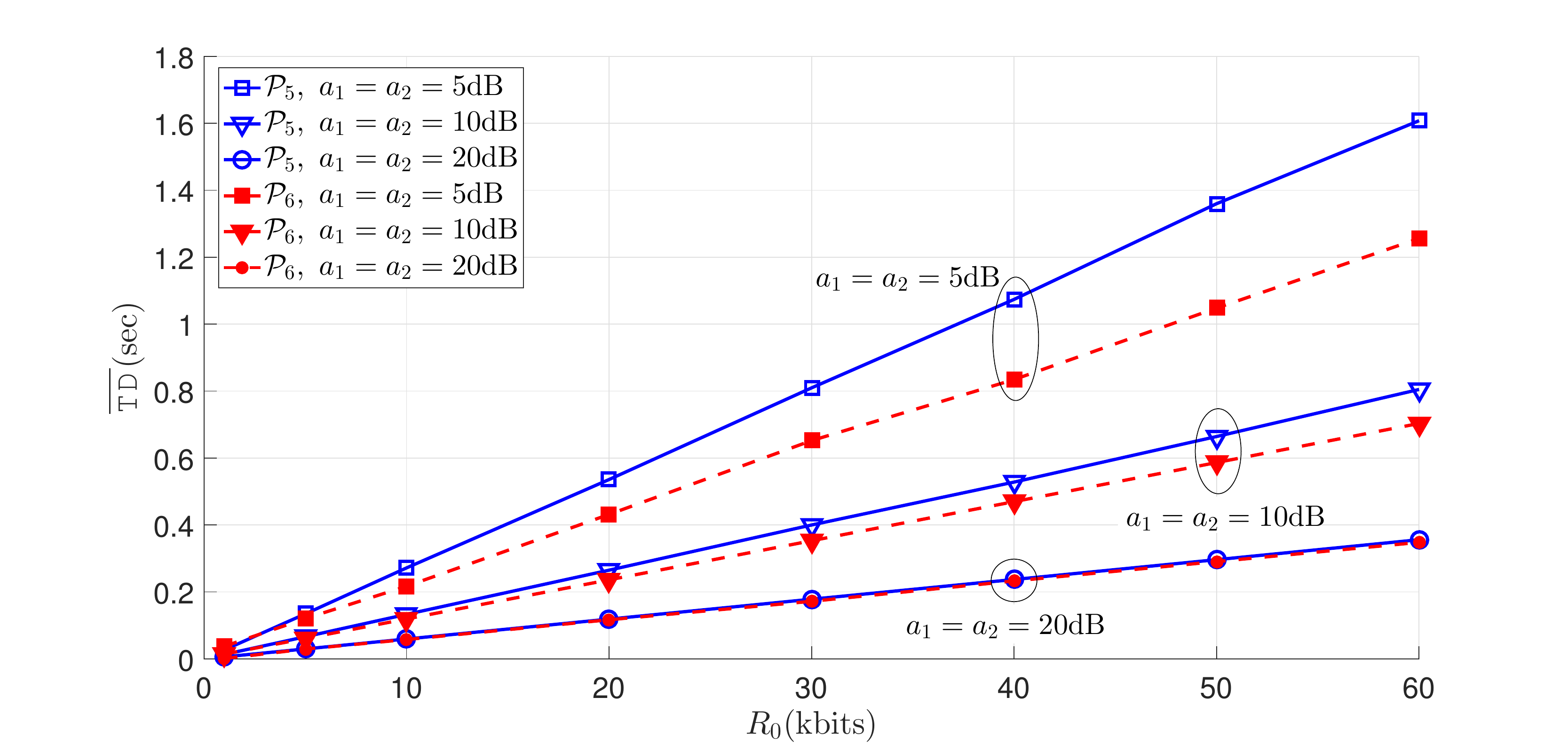}
	\end{center}
	\caption{Comparison of $\overline{\textsf{TD}}$ in different SNRs in the multiuser case when $K=2$.}
	\label{fig:6}
\end{figure}

In the multiuser scenario, we assume $K=2$ and plot the average TD for the problems $\mathcal{P}_5$ to $\mathcal{P}_6$ versus $R_0$ in Fig. \ref{fig:6}. Here, $m=4$, $a_1=a_2$ and we compare the results for three average SNRs $a_1=a_2=5\textrm{dB}$, $10\textrm{dB}$  and $20\textrm{dB}$. Note that in the problem $\mathcal{P}_6$, we have power allocation at the HAP and the obtained results show the benefit of this power allocation in the low SNR. Therefore, in the multiuser case and very similar to the single user counterpart, the optimal power allocation at the HAP is recommended and can improve the average TD especially at the low SNR. 

\section{Conclusion}\label{sec:conclusion}
In this paper we study transmission delay minimization in a point-to-point system with one HAP and one WPC node. For this reason, we first define four optimization problems $\mathcal{P}_1$ to $\mathcal{P}_4$ with different conditions. In the problems $\mathcal{P}_1$ and $\mathcal{P}_2$, we assume an equal DL and UL time duration and in the problems $\mathcal{P}_3$ and $\mathcal{P}_4$, we assume an unequal DL and UL time duration which can be optimally allocated to the HAP and WPC node for the power transmission and data transmission respectively. In addition, an optimal power allocation is assumed for the HAP in the problems $\mathcal{P}_2$ and $\mathcal{P}_4$ as well. Then, the optimal allocated time duration and transmit power are derived. We express that the optimal solutions can be used in the \nakagami fading channel with $m>2$ and hence, use of these techniques are not possible in Rayleigh fading channel. In addition, we compare the results and show that the simultaneous time and power allocation in the problem $\mathcal{P}_4$ leads to the lowest transmission delay among the other methods. However, extraction of this solution has high complexity and therefore, a simple and tight approximation for the solution is presented. After that, the optimization problems are extended to the multiuser scenario and we define two different minimization problems $\mathcal{P}_5$ and $\mathcal{P}_6$. In these two problems, minimization of total TD between one HAP and $K$ nodes is considered with and without the optimal power allocation at the HAP. Finally we conclude that the optimal time and power allocation can be quite effective for a WPC system in low and moderate SNRs.


\appendix
\section{Convergence of \eqref{eq:mean_rtt_opt_p1}}\label{apndx:A}
\renewcommand{\theequation}{\Alph{section}-\arabic{equation}}
\setcounter{equation}{0}
We introduce a continuous and differentiable function $z_1(x)$ over $(0,\infty)$ as
\begin{equation}\label{eq:apndx_a_1}
z_1(x) = \underbrace{\dfrac{2R_0m^m}{B\ln2\Gamma(m)}}_{c_1}\dfrac{x^{m-1}\e^{-mx}}{\ln(1+a_0x^2)}
\end{equation}
with $a_0>0,~m>0$ and redefine \eqref{eq:mean_rtt_opt_p1} as 
\begin{equation}\label{eq:apndx_a_2}
\overline{\textsf{TD}}=\int_0^{\infty}z_1(x)dx=\underbrace{\int_0^1 z_1(x)dx}_{\overline{\textsf{TD}}_1}+\underbrace{\int_1^{\infty}z_1(x)dx}_{\overline{\textsf{TD}}_2}
\end{equation}
and we prove that for $m>2$, $\overline{\textsf{TD}}_1$ and $\overline{\textsf{TD}}_2$ are convergent and so does $\overline{\textsf{TD}}$.

For convergence testing of $\overline{\textsf{TD}}_1$, we choose the comparison function $p_1(x)$ as
\begin{equation}\label{eq:apndx_a_3}
p_1(x)=\dfrac{x^{m-1}}{a_0x^2}=\dfrac{1}{a_0x^{3-m}}
\end{equation}
such that
\begin{equation}\label{eq:apndx_a_4}
\lim_{x\to 0}\dfrac{z_1(x)}{p_1(x)}=c_1.
\end{equation}
The limit comparison test \cite{calculus} says that $\overline{\textsf{TD}}_1$ converges if and only if $\int_0^1p_1(x)dx$ converges. We now use the direct comparison test (some times called p-test) for the convergence of $\int_0^1p_1(x)dx$, which says that $\int_0^1p_1(x)dx$ is a type II improper integral and converges if and only if $3-m<1$. Therefore, choosing $m>2$ leads to the convergence of $\overline{\textsf{TD}}_1$ as well.

Then, for testing $\overline{\textsf{TD}}_2$, from the Taylor expansion of $\e^{mx}$ at large $x$, we know that
\begin{equation}\label{eq:apndx_a_5}
\e^{mx}>\frac{(mx)^{m+1}}{(m+1)!}.
\end{equation}
So, for large $x$ we can write 
\begin{equation}\label{eq:apndx_a_6}
z_1(x)<\underbrace{\dfrac{c_1(m+1)!}{m^{m+1}}}_{c^\prime_1}\dfrac{x^{m-1}}{x^{m+1}\ln(1+a_0x^2)}<\underbrace{c^\prime_1\dfrac{1}{x^2}}_{q_1(x)}.
\end{equation}
$\int_{1}^{\infty}q_1(x)$ is a type I improper integral which is always convergent. Therefore, $\overline{\textsf{TD}}_2$ is convergent too. 

Finally, we can conclude that choosing $m>2$ leads to the convergence of $\overline{\textsf{TD}}_1$ and $\overline{\textsf{TD}}_2$ and consequently $\overline{\textsf{TD}}$.

\section{Derivation of \eqref{eq:solution_beta_opt_p2} and \eqref{eq:solution_T0_opt_p2}}\label{apndx:B}
\renewcommand{\theequation}{\Alph{section}-\arabic{equation}}
\setcounter{equation}{0}
Using Lagrangian method, the cost function is written as
\begin{equation}\label{eq:apndx_b_1}
J=-2T_0+\lambda\left(-R_0+BT_0\log_2 \left(1+\beta a_0\h^2\right)\right)+\mu\left(1-\mathbb{E}\left\{\beta\right\}\right)
\end{equation}
where $\lambda\geq 0$, $\mu\geq 0$ represent the Lagrange multipliers corresponding to the constraints \eqref{eq:const_opt_p2_a} and \eqref{eq:const_opt_p2_b} respectively. Taking the partials with respect to $T_0$ and $\beta$, we obtain
\begin{equation}\label{eq:apndx_b_2}
\dfrac{\partial J}{\partial T_0}=-2+\lambda B\log_2\left(1+\beta a_0\h^2\right)
\end{equation}
and
\begin{equation}\label{eq:apndx_b_3}
\dfrac{\partial J}{\partial \beta}=\dfrac{\lambda B}{\ln2}\dfrac{T_0a_0\h^2}{1+\beta a_0\h^2}-\mu.
\end{equation}
Now, we have to find $T_0^*$, $\beta^*$, $\lambda^*$ and $\mu^*$ such that
\begin{subequations}\label{eq:apndx_b_4}
	\begin{align}
	\label{eq:apndx_b_4_a}
	&-2+\lambda^* B\log_2\left(1+\beta^* a_0\h^2\right)=0
	\\                                                                      
	\label{eq:apndx_b_4_b}
	&\dfrac{\lambda^* B}{\ln2}\dfrac{T_0^*a_0\h^2}{1+\beta^* a_0\h^2}-\mu^*=0
	\\
	\label{eq:apndx_b_4_c}
	&\lambda^*\left(-R_0+B T_0^*\log_2\left(1+\beta^*a_0\h^2\right)\right)=0
	\\
	\label{eq:apndx_b_4_d}
	&\mu^*\left(1-\mathbb{E}\left\{\beta^*\right\}\right)=0.
	\end{align}
\end{subequations}
From \eqref{eq:apndx_b_4_a}, we can conclude that $\lambda^*\neq 0$ and then from \eqref{eq:apndx_b_4_b}, we find that $\mu^*\neq 0$. Therefore, we can rewrite \eqref{eq:apndx_b_4} as
\begin{subequations}\label{eq:apndx_b_5}
	\begin{align}
	\label{eq:apndx_b_5_a}
	&\lambda^* B\log_2\left(1+\beta^* a_0\h^2\right)=2
	\\                                                                      
	\label{eq:apndx_b_5_b}
	&\dfrac{\lambda^* B}{\ln2}\dfrac{T_0^*a_0\h^2}{1+\beta^* a_0\h^2}=\mu^*
	\\
	\label{eq:apndx_b_5_c}
	&B T_0^*\log_2\left(1+\beta^*a_0\h^2\right)=R_0
	\\
	\label{eq:apndx_b_5_d}
	&\mathbb{E}\left\{\beta^*\right\}=1.
	\end{align}
\end{subequations}
From \eqref{eq:apndx_b_5_a}, we write $\lambda^*$ in term of $\beta^*$ and from \eqref{eq:apndx_b_5_b} we write $T_0^*$ in term of $\beta^*$ and insert them into \eqref{eq:apndx_b_5_c} to obtain
\begin{equation}\label{eq:apndx_b_6}
\left(1+\beta^*a_0\h^2\right)\ln^2\left(1+\beta^*a_0\h^2\right)=\dfrac{2\ln2a_0\h^2R_0}{B\mu^*}.
\end{equation}
Finally, \eqref{eq:apndx_b_6} can be solved according to
\begin{equation}\label{eq:apndx_b_7}
\beta^*=\dfrac{\e^{2\mathcal{W}_0\left(\sqrt{\dfrac{2\ln2a_0\h^2R_0}{B\mu^*}}\Bigg/2\right)}-1}{a_0\h^2}
\end{equation}
where $\mathcal{W}_0(.)$ represents Lambert-W function \cite{lambertw}. Then, by inserting \eqref{eq:apndx_b_7} into \eqref{eq:apndx_b_5_c}, $T_0^*$ is attained and $\mu^*$ is calculated from \eqref{eq:apndx_b_5_d}.

\section{Convergence of \eqref{eq:unity_power_opt_p2} and  \eqref{eq:mean_rtt_opt_p2}}\label{apndx:C}
\renewcommand{\theequation}{\Alph{section}-\arabic{equation}}
\setcounter{equation}{0}
The proof is similar to \ref{apndx:A} and we explain the process for both \eqref{eq:mean_rtt_opt_p2} and \eqref{eq:unity_power_opt_p2} respectively. First we use $z_2(x)$
\begin{equation}\label{eq:apndx_c_1}
z_2(x)=\underbrace{\dfrac{R_0\ln2m^m}{B\Gamma(m)}}_{c_{21}}\dfrac{x^{m-1}\e^{-mx}}{\mathcal{W}_0\left(\underbrace{\sqrt{\dfrac{\ln2a_0R_0}{2B\mu^*}}}_{c_{22}}~x\right)}=c_{21}\dfrac{x^{m-1}\e^{-mx}}{\mathcal{W}_0\left(c_{22}x\right)}
\end{equation}
over $(0,\infty)$ with $c_{21}>0,~c_{22}>0$ and redefine \eqref{eq:mean_rtt_opt_p2} as 
\begin{equation}\label{eq:apndx_c_2}
\overline{\textsf{TD}}=\int_0^{\infty}z_2(x)dx=\underbrace{\int_0^1 z_2(x)dx}_{\overline{\textsf{TD}}_1}+\underbrace{\int_1^{\infty}z_2(x)dx}_{\overline{\textsf{TD}}_2}.
\end{equation}
Regarding the Taylor expansion of $\mathcal{W}_0(x)$ near $0$ as \cite{lambertw}
\begin{equation}\label{eq:apndx_c_3}
\mathcal{W}_0(x)=\sum_{n=1}^{\infty}\dfrac{(-n)^{n-1}}{n!}x^n,
\end{equation}
and in a quite similar way to the \ref{apndx:A}, we can find that $\overline{\textsf{TD}}$ in \eqref{eq:mean_rtt_opt_p2} is convergent when $m>1$.

Now, we go on to the proof of convergence for \eqref{eq:unity_power_opt_p2} and introduce $z^\prime_2(x)$ as
\begin{eqnarray}\label{eq:apndx_c_4}
z^{\prime}_2(x)&=&\underbrace{\dfrac{m^m}{\Gamma(m)}}_{c_{23}}\left(\dfrac{\e^{2\mathcal{W}_0(c_{22}x)}x^{m-1}\e^{-mx}}{a_0x^2}-\dfrac{x^{m-1}\e^{-mx}}{a_0x^2}\right)\nonumber\\
&=&\underbrace{c_{23}\dfrac{\e^{2\mathcal{W}_0(c_{22}x)}x^{m-1}\e^{-mx}}{a_0x^2}}_{z^\prime_{21}(x)}-\underbrace{c_{23}\dfrac{x^{m-1}\e^{-mx}}{a_0x^2}}_{z^\prime_{22}(x)},
\end{eqnarray}
then redefine \eqref{eq:unity_power_opt_p2} as
\begin{equation}\label{eq:apndx_c_5}
\mathbb{E}\{\beta^*\}=\underbrace{\int_0^{1}z^\prime_{21}(x)dx}_{I_1}+\underbrace{\int_1^{\infty}z^\prime_{21}(x)dx}_{I_2}-\underbrace{\int_0^{1}z^\prime_{22}(x)dx}_{I_3}-\underbrace{\int_1^{\infty}z^\prime_{22}(x)dx}_{I_4}.
\end{equation}
$I_1$ is a type II improper itegral and we can choose comparison function $p^{\prime}_{21}(x)=\dfrac{x^{m-1}}{a_0x^2}=\dfrac{1}{a_0x^{3-m}}$ such that
\begin{equation}\label{eq:apndx_c_6}
\lim_{x\to 0}\dfrac{z^{\prime}_{21}(x)}{p^{\prime}_{21}(x)}=c_{23},
\end{equation}
so, using the limit comparison test \cite{calculus}, we find that $I_1$ is convergent when $m>2$. Again, $I_2$ is a type I improper integral and we know that for large $x$, $\e^{\mathcal{W}_0(x)}<x$ \cite{lambertw}. Therefore,
\begin{equation}\label{eq:apndx_c_7}
z^{\prime}_{21}(x)<c_{23}\dfrac{2c_{22}xx^{m-1}\e^{-mx}}{a_0x^2},
\end{equation}
So, $I_2$ is always convergent and does not depend on $m$. In a similar way, $I_3$ is a type II improper integral and it is convergent for $m>2$ and $I_4$ is a type I improper integral, but always convergent. 

Finally, we can conclude that choosing $m>2$, leads to the convergence of both \eqref{eq:unity_power_opt_p2} and  \eqref{eq:mean_rtt_opt_p2}. 

\section{Derivation of \eqref{eq:solution_beta_opt_p4},  \eqref{eq:solution_T1_opt_p4} and \eqref{eq:solution_T2_opt_p4}}\label{apndx:D}
\renewcommand{\theequation}{\Alph{section}-\arabic{equation}}
\setcounter{equation}{0}
The cost function can be written as
\begin{equation}\label{eq:apndx_d_1}
J=-(T_1+T_2)+\lambda\left(-R_0+BT_2\log_2 \left(1+\beta a_0\h^2\dfrac{T_1}{T_2}\right)\right)+\mu\left(1-\mathbb{E}\left\{\beta\right\}\right)
\end{equation}
where $\lambda\geq 0$, $\mu\geq 0$ represent the Lagrange multipliers corresponding to the constraints \eqref{eq:const_opt_p4_a} and \eqref{eq:const_opt_p4_b} respectively. Taking the partials with respect to $T_1$, $T_2$ and $\beta$, we will have
\begin{equation}\label{eq:apndx_d_2}
\dfrac{\partial J}{\partial T_1}=-1+\lambda\dfrac{B}{\ln2}\dfrac{\beta a_0\h^2}{1+\beta a_0\h^2\dfrac{T_1}{T_2}}
\end{equation}
\begin{equation}\label{eq:apndx_d_3}
\dfrac{\partial J}{\partial T_2}=-1+\lambda\left(B\log_2\left(1+\beta a_0\h^2\dfrac{T_1}{T_2}\right)-\dfrac{B}{\ln2}\dfrac{\beta a_0\h^2\dfrac{T_1}{T_2}}{1+\beta a_0\h^2\dfrac{T_1}{T_2}}\right)
\end{equation}
and
\begin{equation}\label{eq:apndx_d_4}
\dfrac{\partial J}{\partial \beta}=\dfrac{\lambda B}{\ln2}T_2\dfrac{a_0\h^2\dfrac{T_1}{T_2}}{1+\beta a_0\h^2\dfrac{T_1}{T_2}}-\mu.
\end{equation}
Now, we have to find $T_1^*$, $T_2^*$, $\beta^*$, $\lambda^*$ and $\mu^*$ such that
\begin{subequations}\label{eq:apndx_d_5}
	\begin{align}
	\label{eq:apndx_d_5_a}
	&-1+\lambda^*\dfrac{B}{\ln2}\dfrac{\beta^*a_0\h^2}{1+\beta^*a_0\h^2\dfrac{T_1^*}{T_2^*}}=0
	\\                                                                      
	\label{eq:apndx_d_5_b}
	&-1+\lambda^*\left(B\log_2\left(1+\beta^* a_0\h^2\dfrac{T_1^*}{T_2^*}\right)-\dfrac{B}{\ln2}\dfrac{\beta^*a_0\h^2\dfrac{T_1^*}{T_2^*}}{1+\beta^*a_0\h^2\dfrac{T_1^*}{T_2^*}}\right)=0
	\\
	\label{eq:apndx_d_5_c}
	&\dfrac{\lambda^*B}{\ln2}T_2^*\dfrac{a_0\h^2\dfrac{T_1^*}{T_2^*}}{1+\beta^* a_0\h^2\dfrac{T_1^*}{T_2^*}}-\mu^*=0
	\\
	\label{eq:apndx_d_5_d}
	&\lambda^*\left(-R_0+BT_2^*\log_2\left(1+\beta^*a_0\h^2\dfrac{T_1^*}{T_2^*}\right)\right)=0
	\\
	\label{eq:apndx_d_5_e}
	&\mu^*\left(1-\mathbb{E}\left\{\beta^*\right\}\right)=0.
	\end{align}
\end{subequations}
From \eqref{eq:apndx_d_5_a}, we can conclude that $\lambda^*\neq 0$ and then from \eqref{eq:apndx_d_5_c}, we find that $\mu^*\neq 0$. Therefore, we can rewrite \eqref{eq:apndx_d_5} as
\begin{subequations}\label{eq:apndx_d_6}
	\begin{align}
	\label{eq:apndx_d_6_a}
	&\lambda^*\dfrac{B}{\ln2}\dfrac{\beta^*a_0\h^2}{1+\beta^*a_0\h^2\tau^*}=1
	\\                                                                      
	\label{eq:apndx_d_6_b}
	&\lambda^*B\log_2\left(1+\beta^* a_0\h^2\tau^*\right)=1+\lambda^*\dfrac{B}{\ln2}\dfrac{\beta^*a_0\h^2\tau^*}{1+\beta^*a_0\h^2\tau^*}
	\\
	\label{eq:apndx_d_6_c}
	&\lambda^*\dfrac{B}{\ln2}T_2^*\dfrac{a_0\h^2\tau^*}{1+\beta^*a_0\h^2\tau^*}=\mu^*
	\\
	\label{eq:apndx_d_6_d}
	&BT_2^*\log_2\left(1+\beta^*a_0\h^2\tau^*\right)=R_0
	\\
	\label{eq:apndx_d_6_e}
	&\mathbb{E}\left\{\beta^*\right\}=1
	\end{align}
\end{subequations}
where $\tau^*=T_1^*/T_2^*$. From \eqref{eq:apndx_d_6_a}, we write $\lambda^*$ in term of $\beta^*$ and $\tau^*$ to reduce the equations as
\begin{subequations}\label{eq:apndx_d_7}
	\begin{align}
	\label{eq:apndx_d_7_a}
	&\dfrac{1+\beta^*a_0\h^2\tau^*}{\beta^*a_0\h^2}\ln\left(1+\beta^*a_0\h^2\tau^*\right)=1+\tau^*
	\\                                                                      
	\label{eq:apndx_d_7_b}
	&T_2^*\tau^*=\beta^*\mu^*
	\\
	\label{eq:apndx_d_7_c}
	&BT_2^*\log_2\left(1+\beta^*a_0\h^2\tau^*\right)=R_0
	\\
	\label{eq:apndx_d_7_d}
	&\mathbb{E}\left\{\beta^*\right\}=1.
	\end{align}
\end{subequations}
Moreover from \eqref{eq:apndx_d_7_c} we solve $T_2^*$ in term of $\beta^*$ and $\tau^*$. Therefore we will have
\begin{subequations}\label{eq:apndx_d_8}
	\begin{align}
	\label{eq:apndx_d_8_a}
	&\dfrac{1+\beta^*a_0\h^2\tau^*}{\beta^*a_0\h^2}\ln\left(1+\beta^*a_0\h^2\tau^*\right)=1+\tau^*
	\\                                                                      
	\label{eq:apndx_d_8_b}
	&\dfrac{R_0\ln2}{B}\tau^*=\beta^*\mu^*\ln\left(1+\beta^*a_0\h^2\tau^*\right)
	\\
	\label{eq:apndx_d_8_c}
	&\mathbb{E}\left\{\beta^*\right\}=1.
	\end{align}
\end{subequations}
Next, from \eqref{eq:apndx_d_8_a} and Lambert-W function definition \cite{lambertw}, we can write 
\begin{equation}\label{eq:apndx_d_9}
1+\beta^*a_0\h^2\tau^*=\e^{1+\mathcal{W}_0\left(\dfrac{\beta^*a_0\h^2-1}{\e}\right)}
\end{equation}
and insert it into \eqref{eq:apndx_d_8_b} to obtain \eqref{eq:solution_beta_opt_p4}. In addition, $\mu^*$ is calculated from \eqref{eq:apndx_d_8_c}. Then, $\tau^*$ is derived from \eqref{eq:apndx_d_9} as
\begin{equation}\label{eq:apndx_d_10}
\tau^*=\dfrac{\e^{1+\mathcal{W}_0\left(\dfrac{\beta^*a_0\h^2-1}{\e}\right)}-1}{\beta^*a_0\h^2}
\end{equation}
and $T_2^*$ is obtained from \eqref{eq:apndx_d_7_b} as
\begin{equation}\label{eq:apndx_d_11}
T_2^*=\dfrac{\beta^*\mu^*}{\tau^*}.
\end{equation}
Now, we can use \eqref{eq:apndx_d_10} and \eqref{eq:apndx_d_11} to extract \eqref{eq:solution_T1_opt_p4} and \eqref{eq:solution_T2_opt_p4} respectively.

\bibliographystyle{unsrt}
\bibliography{References}

\end{document}